\documentclass[useAMS,usenatbib,usegraphicx]{mn2e}
\usepackage{times}

\newcommand{\lsim}{\lower0.6ex\vbox{\hbox{$ \buildrel{\textstyle <}\over{\sim}\ $}}}
\newcommand{\gsim}{\lower0.6ex\vbox{\hbox{$ \buildrel{\textstyle >}\over{\sim}\ $}}}
\newcommand{\lesssim}{\lower0.6ex\vbox{\hbox{$ \buildrel{\textstyle <}\over{\sim}\ $}}}
\newcommand{\gtrsim}{\lower0.6ex\vbox{\hbox{$ \buildrel{\textstyle >}\over{\sim}\ $}}}
\newcommand{\beq}{\begin{equation}}
\newcommand{\eeq}{\end{equation}}
\title[Cosmological Simulations of Decaying Dark Matter]
{
Cosmological Simulations of Decaying Dark Matter: Implications for Small-scale Structure of Dark Matter Halos
}

\author[M.-Y. Wang et al.]
{Mei-Yu Wang$^{1,2}$\thanks{Email:meiywang@indiana.edu}, Annika H. G. Peter$^{3,4}$, Louis E. Strigari$^1$, Andrew R. Zentner$^2$,
\newauthor
, Bryan Arant$^4$, Shea Garrison-Kimmel$^4$, and Miguel Rocha$^4$ \\
$^1$Department of Physics, Indiana University--Bloomington, Bloomington, IN 47405-7105\\
$^2$Department of Physics $\&$ Astronomy, and Pittsburgh Particle physics, Astrophysics, and Cosmology Center (Pitt PACC), 
University of Pittsburgh, Pittsburgh, PA 15260 \\
$^3$CCAPP, Department of Physics, and Department of Astronomy, The Ohio State University, Columbus, OH 43210\\
$^4$Department of Physics and Astronomy, University of California, Irvine, CA 92697-4575
}
\date{Released 2014 Xxxxx XX}

\pagerange{\pageref{firstpage}--\pageref{lastpage}} \pubyear{2014}

\begin{document}

\maketitle

%----------------------------------------------------------------
%%%%%%%%%%%%%%%%%%%%%%%  A B S T R A C T %%%%%%%%%%%%%%%%%%%%%%%%%%%%%%

\begin{abstract}

We present a set of N-body simulations of a class of models in which an unstable dark matter particle decays into a stable non-interacting dark matter particle, with decay lifetime comparable to the Hubble time. We study the effects of the kinematic recoil velocity ($V_k$) received by the stable dark matter on the structures of dark matter halos ranging from galaxy-cluster to Milky Way mass scales. For Milky Way-mass halos, we use high-resolution, zoom-in simulations to explore the effects of decays on Galactic substructure. In general, halos with circular velocities comparable to the magnitude of kick velocity are most strongly affected by decays. We show that decaying dark matter models with lifetimes $\Gamma^{-1}$ $\sim H_0^{-1}$ and recoil speeds $V_k \sim 20-40\,\mathrm{km/s}$ can significantly reduce both the abundance of Galactic subhalos and the internal densities of the subhalos. We also compare subhalo circular velocity profiles with observational constraints on the Milky Way dwarf satellite galaxies. Interestingly, we find that decaying dark matter models that do not violate current astrophysical constraints, can significantly mitigate {\em both} the well-documented ``missing satellites problem" and the more recent ``too big to fail problem" associated with the abundances and densities of Local Group dwarf satellite galaxies. A relatively unique feature of late decaying dark matter models is that they predict significant evolution of halos as a function of time. This is an important consideration because at high redshifts, prior to decays, decaying models exhibit the same sequence of structure formation as cold dark matter.  Thus, decaying dark matter models are significantly less constrained by high-redshift phenomena (e.g., reionization, AGN formation, Lyman-$\alpha$ forest) than warm mark matter models that exhibit similar low-redshift predictions. We conclude that models of decaying dark matter make predictions that are relevant for the interpretation of observations of small galaxies in the Local Group and can be tested or constrained by the kinematics of Local Group dwarf galaxies as well as by forthcoming large-scale surveys.  
%We compare subhalo circular velocity profiles of this model to the corresponding predictions of cold, warm, and self-interacting dark matter models, highlighting model differences and similarities.
\end{abstract} 
  
%---------------------------
\section{Introduction} 
\label{section:introduction} 
%---------------------------
%
The formation of structure in the universe is driven by dark matter (DM), the nature of which remains unknown. Over the last several decades, the hierarchical cold dark matter model (CDM) has become the standard description for the formation of cosmic structure and galaxy formation \citep{White_etal78,Blumenthal_etal84}. In particular, the CDM model is consistent with the cosmic microwave background (CMB) anisotropy spectrum measured by the Wilkinson Microwave Anisotropy Probe (WMAP) \citep{WMAP9} and PLANCK satellite \citep{Planck_13} as well as observations of the large-scale (k $\lsim$ 0.1 $h$/Mpc) galaxy clustering spectrum measured by the Sloan Digital Sky Survey (SDSS) \citep{Tegmark_etal06}. However, the CDM paradigm faces a number of possible challenges on small scales. For example, Galaxy-sized CDM halos contain a large number of subhalos in numerical simulations~\citep{klypin_etal99b,moore_etal99}, while observations show only $\sim 20$ satellite galaxies around each of the Milky Way and M31~\citep{Koposov_etal08}. In addition, the central slopes of the density profiles of low-surface brightness (LSB) galaxies~\citep{deBlok_etal02,Simon_etal05,KuziodeNaray_etal08,Oh_etal11,Adams2014} appear less than than predicted in CDM simulations. 

A number of proposals have been made to reconcile the observations with the theory. First and foremost, baryonic physics renders galaxy formation inefficient in small halos ($M \lsim 10^9 \mathrm{M}_\odot$).  Thus, many subhalos may exist in Galactic-scale halos but may not be inhabited by galaxies \citep{bullock2000,benson2002,Sawala:2014baa}.   This would leave small sub halos invisible to optical surveys.  To address the question of central densities, on somewhat larger scales ($M \lsim 10^{11} \mathrm{M}_{\odot}$), baryonic processes may significantly alter the distribution of dark matter within small dark matter halos. Several groups performed high-resolution hydrodynamic simulations with stellar or supernovae feedback, showing that it may be possible to generate shallow dark matter profiles if feedback is sufficiently strong~\citep{Governato_etal12,Teyssier_etal13,Garrison-Kimmel_etal13}.  For isolated galaxies, it appears that the generation of a shallow density profile similar to that observed in LSBs cored profile may require fairly large stellar masses ($M_*\sim 10^8 M_{\odot}$). Most low luminosity dwarf spheroidal (dSph) galaxies ($M_*\sim 10^6 M_{\odot}$) in our Local Group are likely not affected by this mechanism~\citep{Governato_etal12,Garrison-Kimmel_etal13}.  For satellite galaxies, however, \citet{brooks_etal13} argued that strong feedback combined with enhanced tidal stripping of less-concentrated satellite galaxies may alleviate the apparent discrepancies that arise on the scale of the dwarf galaxies of the MW. 

Though baryonic resolutions to the small-scale challenges to CDM must be explored exhaustively prior to reaching strong conclusions about the dark matter, it is also worthwhile to consider the predictions of alternative dark matter physics. Altering the cold or collisionless properties of the dark matter may result in shallow density profiles and fewer Milky Way satellites, while maintaining the large scale successes of the LCDM paradigm. Recently, several groups have used high-resolution simulations to study how alternative dark matter models alter halo properties. Some thermal relic warm dark matter (WDM) candidates, such as sterile neutrinos and gravitinos, have masses of $\sim$ few keV and velocities at production substantial enough to suppress structure growth on Galactic and sub-Galactic scales \citep{Polisensky_etal11, Maccio_etal12,Lovell_etal12,Lovell_etal14,Horiuchi2014,Schneider2014}. In these scenarios, Galactic subhalos tend to be less concentrated, because a halo at a fixed mass scale forms later in a WDM Universe than in a CDM Universe. WDM can also be produced in the early universe through {\it resonant oscillations}~\citep{Shi:1998km} and such scenarios often yield a mixture of both cold and warm DM models (CWDM). CWDM also results in suppressed small-scale structure growth~\citep{Maccio_etal12b,Anderhalden_etal13}. WDM models are now somewhat strongly constrained by Lyman-$\alpha$ forest data, with a lower mass bound of $m_{\mathrm{WDM}} \gsim 3\, \mathrm{keV}$ \citep[thermal relic mass;][]{Viel_etal13}. This severely limits the ability of WDM to make both shallow DM density profiles, and a population of MW satellites similar to that which is observed \citep{Polisensky_etal11,Villaescusa-Navarro_etal11,Maccio_etal12}. However, CWDM models, which have non-zero power below free-streaming scale~\citep{Maccio_etal12b}, remain viable. In addition to WDM and CWDM, self-interacting dark matter (SIDM) models~\citep{Zurek:2013wia}, both with velocity-independent~\citep{Rocha_etal12} and velocity-dependent cross sections~\citep{Vogelsberger_etal12}, have been studied in the context of structure formation. These authors find that the inner density profiles of halos can be made significantly shallower and less dense compared to CDM because of DM self-scattering. Recently there are also a few works on DM interacting with relativistic species \citep{Boehm_etal14,Buckley_etal14}.
%
%--------------------------------------------------------------------
%
\begin{table*}
\caption{Numerical Parameters of Cosmological Simulations}
{\renewcommand{\arraystretch}{1.3}
\renewcommand{\tabcolsep}{0.2cm}
\begin{tabular}{l c c c c c c c}
\hline 
\hline
Name & Box size & Number of Particles & Particle mass & Force Softening & Decay Lifetime & Kick Velocity\\
  &[$h^{-1}$ Mpc] &  $N_p$ &  $m_p$ [$M_{\odot}$] &  $\epsilon$ [pc] & $\Gamma^{-1}$ [Gyr]& $V_{k}$ [km/s]\\
\hline
B50-CDM &50&$512^3$ &9.83$\times 10^7$ & 1400 &- &-\\
B50-t40-v100 &50&$512^3$ &9.83$\times 10^7$ & 1400&40 &100.0\\
\hline
Z12-CDM &(3$R_{\mathrm{vir}})^*$ &5.6$\times 10^7$ &1.92$\times 10^5$ & 143 &- &-\\
Z12-t40-v100 &(3$R_{\mathrm{vir}})^*$&5.6$\times 10^7$ &1.92$\times 10^5$ & 143&40 &100.0\\
Z12-t40-v40 &(3$R_{\mathrm{vir}})^*$&5.6$\times 10^7$ &1.92$\times 10^5$ & 143&40 &40.0\\
Z12-t40-v20 &(3$R_{\mathrm{vir}})^*$&5.6$\times 10^7$ &1.92$\times 10^5$ & 143&40 &20.0\\
Z12-t20-v20 &(3$R_{\mathrm{vir}})^*$&5.6$\times 10^7$ &1.92$\times 10^5$ & 143&20 &20.0\\
Z12-t10-v20 &(3$R_{\mathrm{vir}})^*$&5.6$\times 10^7$ &1.92$\times 10^5$ & 143&10 &20.0\\
Z12-t1-v20 &(3$R_{\mathrm{vir}})^*$&5.6$\times 10^7$ &1.92$\times 10^5$ & 143&1 &20.0\\
\hline
Z13-CDM &(3$R_{\mathrm{vir}})^*$ &4.4$\times 10^8$ &2.40$\times 10^4$ & 72 &- &-\\
Z13-t10-v20 &(3$R_{\mathrm{vir}})^*$ &4.4$\times 10^8$ &2.40$\times 10^4$ & 72 &10 &20.0\\
\end{tabular}
\medskip
\\
$*$The volumes listed for the Z12 $\&$ Z13 zoom-in simulations refer to the number of virial radii used to find the Lagrangian volumes associated with the zoom. The virial radius here refers to the CDM halo radius, which is $R_{\mathrm{vir}}$ = 265 kpc. The particle properties listed are for the highest resolution particles only. 
 }
\end{table*}
%
%---------------------------------------------------------------------------------------------------------
%

In this paper, we utilize high-resolution simulations of cosmological structure formation to study the implications of decaying dark matter (DDM) on halo structure and substructure. These types of DDM models have been considered in a number of recent studies~\citep{Strigari:2006jf,peter_10,peter_benson10,peter_etal10, wang_zentner10,wang_zentner12,Wang_etal13}. In such models, a DM particle of mass $M$ decays into a less massive daughter particle of mass $m=(1-f)M$ and a significantly lighter, relativistic particle, with a lifetime on the order of the age of the Universe. The stable daughter particle acquires a recoil kick velocity the magnitude of which depends upon the mass splitting. The evolution of DDM perturbations with arbitrary mass splitting has been examined by \citet{Aoyama_etal14}. For $f \ll 1$, the kick velocity, $V_k \sim f\, c$, is non-relativistic. As a result of the decays, the DM consists of a mixture of cold and warm components, and is similar in some respects to the CWDM produced via resonant oscillation. The linear matter power spectrum in this class of DDM~\citep{wang_zentner12} is characterized by a step and a plateau on small scales, similar to the power spectrum in CWDM~\citep{boyarsky_etal08}, but its evolution is more significant at late times due to the continuing decay process. 

Despite the many similar features between CWDM and DDM, there are distinct phenomenological advantages to DDM. First, it is a natural way to generate a mixture of cold and warm components, provided the decay lifetime is comparable to or longer than the age of the universe (note that DDM will behave like WDM if the decay lifetime is much shorter than a Hubble time \citep{Kaplinghat_05}). Second, the DDM only alters structure formation in the late universe, when decays are prevalent. CWDM/WDM and DDM are distinct primarily because in CWDM/WDM scenarios the mixture of cold and warm components is established at the time of DM production in the early universe.  The relative abundances of cold and warm species is fixed throughout structure formation. In contrast, the relative abundances of the two components in DDM models changes with time due to the decays.  Thus, the effects of DDM on structure growth become more pronounced at late times due to the increasing relative contribution of the warm component. This is an important feature of DDM models because it {\em enables DDM to evade constraints from early structure growth} \cite[e.g.,][]{yue_chen12,Viel_etal13,schultz_etal14}, which severely restrict WDM models, {\em while simultaneously suppressing sub-Galactic structure in a manner similar to WDM and CWDM}. 

This feature can be seen in studies of the Lyman-$\alpha$ forest in DDM models~\citep{Wang_etal13}. The Lyman-$\alpha$ forest places restrictive bounds on the DDM mass splitting for $\Gamma^{-1}$ $\sim$ a few Gyr, but the constraint loosens quickly as $\Gamma^{-1}$ increases. For relatively short lifetimes, Lyman-$\alpha$ forest data provide the best model-independent constraints on DDM, showing that the parameter space $\Gamma^{-1}$ $\lesssim$ a few Gyr and $V_k$ $\gtrsim$ 40 km/s is excluded. Meanwhile, isolated numerical simulations and semi-analytical models have shown that DDM with kick velocities $V_k \lsim 100 \, \mathrm{km/s}$ and lifetimes $\Gamma^{-1}$ $\lsim$ 30-40 Gyr can have important effects on Galactic substructure~\citep{peter_benson10,peter_etal10}. This implies that there is a large and viable DDM parameter space that has a substantial impact on galactic substructure that has not yet been explored in detail.

In this paper, we run the first set of cosmological DDM simulations for a range of lifetimes and kick velocities, and study the phenomenological implications of DDM models over a range of scales. Throughout the paper, we do not refer to any specific beyond-standard-model physics for late decays \citep[e.g.,][]{Cembranos:2005us}, focusing on the astrophysical effects of such models. We compare our results to other simulations of alternative DM models, such as SIDM and WDM, and to the abundance and structure of the MW dwarf satellite galaxies.

The outline of the paper is as follows. In~\S~\ref{section:methods}, we briefly describe the DDM models. \S~\ref{subsection:decay_models} gives an overview of DDM and establishes our notation. In \S~\ref{subsection:simulations} we describe our CDM and DDM simulations. We present our simulation results in~\S~\ref{section:results}. First we give a picture of the overall statistical effects on large scales from our large volume simulations in~\S~\ref{subsection:LSS}. In~\S~\ref{section:halo structure} we focus on individual halos and study how halo properties change across mass scales of clusters to galactic substructures. In~\S~\ref{subsubsection:halo_abundace} we study subhalo abundance and its dependence upon the parameters of DDM models in general, and then we focus on Galactic halos in~\S~\ref{subsubsection:MW_halo_abundace}. In~\S~\ref{subsubsection:density_profiles} we show effects of DDM on halo density profiles. We focus on Galactic subhalo properties in~\S~\ref{subsection:subhalos}. We first perform resolution tests in~\S~\ref{subsubsection:convergence} to demonstrate the convergence of our simulations on scales relevant to the satellite galaxy population. In~\S~\ref{subsubsection:density_subhalos}, we then study the subhalo density profile changes and quantify changes in terms of fits to an Einasto profile. In \S~\ref{subsubsection:vmax_rmax}, we compare the internal structures of the satellite halos in our DDM simulations to the observed structural properties of the halos that host the MW satellite galaxies. We conclude in \S~\ref{section:conclusion} with a discussion of the ways in which different alternative DM scenarios alter predictions for the properties of Galactic substructure.

%____________________________________________________________---------------------
\section{Methodology}
\label{section:methods}
\subsection{Decaying Dark Matter Models}
\label{subsection:decay_models}

We begin by establishing our notation. We consider a class of models in which a DM particle decays into another species of stable DM with a small mass splitting, DDM~$\rightarrow$~SDM~$+$~L, where L denotes a light non-interacting daughter particle, SDM is the stable DM with mass $m$, and DDM is the decaying DM with mass $M$. The mass loss fraction $f = (M-m)/M$ of DDM is directly related to the recoil kick velocity deposited to the SDM particle by $f \simeq V_k/c$ from energy-momentum conservation. Therefore there are two independent parameters in this class of DDM models, namely the decay lifetime $\Gamma^{-1}$ (or decay rate $\Gamma$) and the recoil velocity $V_k$ (or mass splitting ratio $f$). As we will describe later, the relevant decay lifetimes in this work are generally large, ranging from a few Gyr to a few times of the age of the universe. We will focus on the case with f $\ll$ 1 for which the $V_k$ is non-relativistic. 

The advantage of the late decay model is that it can have interesting phenomenological consequences on small scales, while preserving the general success of CDM on large scales and at early times. If the decay lifetime is sufficiently short, the behavior of DDM will be similar to WDM with the mass splitting being the analog of the WDM mass. For a large decay lifetime, the key difference between DDM and WDM lies in the evolution of the free-streaming scale as a function of time. While for WDM (and also for standard model neutrinos) the free-streaming scale gradually shrinks as the momenta of the particles redshifts, late decaying DM continuously generates particles with non-negligible peculiar velocities.  This causes the free-streaming scale to grow until late times \citep[see Fig.~3 in ][for a comparison]{wang_zentner12}. We refer readers to~\citet{wang_zentner12} for a detailed discussion and analytical explanations for this class of unstable DM. The full set of perturbation equations for deriving matter perturbations and the evolution of the free-streaming scale is also presented there.

\begin{figure*}
\includegraphics[height=7.0cm]{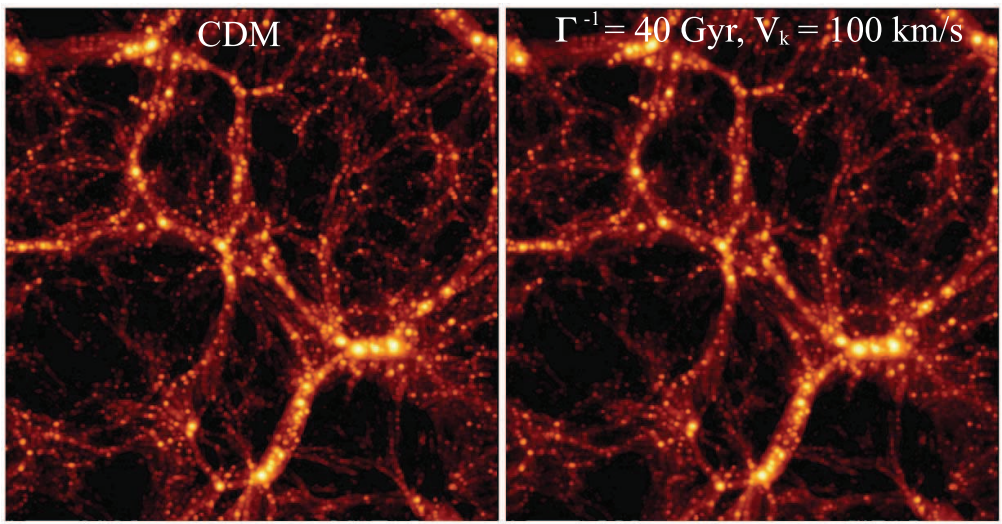}
\caption{ 
Large-scale DM clustering in CDM (left) and DDM with $\Gamma^{-1}$ = 40 Gyr, $V_k$ = 100 km/s (right) of a 10 $h^{-1}$ $Mpc$ deep slice in the 50 $\times$ 50~$h^{-2}$~Mpc$^2$ cosmological box at $z=0$. The color scheme indicates the line-of-sight projected square of the density to emphasize the locations of dense structures, such as halos within filaments. The large-scale structure of the CDM and DDM simulations are virtually identical.
}
\label{fig:snapshot}
\end{figure*}

\begin{figure*}
\includegraphics[height=7.0cm]{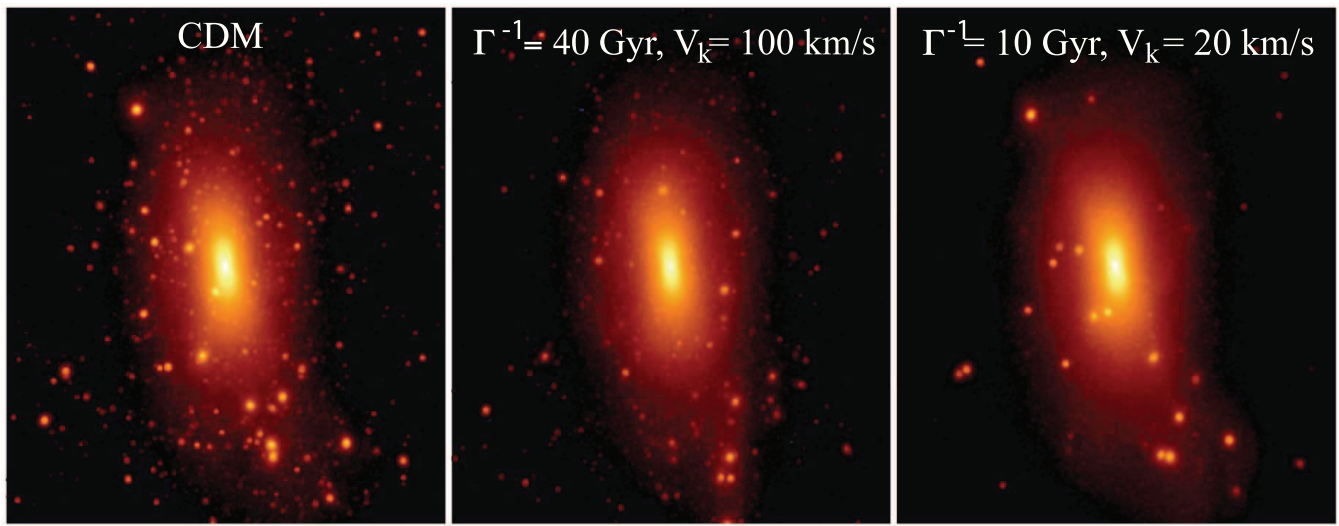}
\caption{ 
Small-scale structure in a Milky Way mass halo (Z12) in CDM (left) and DDM models with $\Gamma^{-1}$ = 40 Gyr and $V_k$ = 100 km/s (middle) and $\Gamma^{-1}$ = 10 Gyr and $V_k$ = 20 km/s (right) within 260 kpc of the halo centers at $z=0$. The color scheme indicates the line-of-sight projected square of the density in order to emphasize the dense structures such as the host halo interiors and the associated subhalos. The DDM halos have slightly more diffuse central regions. The abundance and structure of subhalos are altered significantly compared to CDM in both of the DDM simulations presented.  % {\bf AP: originally this said "halos", but I cannot see any difference between the central regions of the main halos here.} {\bf [Is it possible to label the panels with "CDM", "DDM" and specify the Decay parameters in each panel? That would be helpful. However, I understand if this is difficult or not possible with the software you are using so don't feel absolutely obligated to do this.]} {\bf [MW: It is difficult for me to do at this point]}{\bf AP: You can make an extra row using TeX in the figure and place figure labels either above or below.} {\bf I used photoshops to add the labels.}
}
\label{fig:zoom_sim}
\end{figure*}

\subsection{Simulations}
\label{subsection:simulations}

We have conducted two sets of high-resolution cosmological simulations to study structure formation in DDM models. We performed uniform resolution simulations in a cubic box $50\, h^{-1}\mathrm{Mpc}$ on a side in two models (the "B50" series of simulations). Furthermore, to study the detailed properties of Galaxy-sized halo structure and substructure in DDM models, we performed a suite of zoom simulations \citep{katz_white93,Onorbe_etal14} focused on a halo of mass $M \approx 10^{12}\, \mathrm{M}_{\odot}$ (the "Z12" and "Z13" series). We generated each simulation using a modified version of the parallel N-body code GADGET-2~\citep{Springel_etal05} and GADGET-3 by~\cite{peter_etal10}. The modified version consists of a Monte-Carlo simulation at each time step $\Delta t$ to determine whether a particle should decay with decay probability $P=\Gamma\Delta t$. If a particle is designated for decay, it will receive a kick speed $V_{k}$ in a random direction, and it will be flagged to make sure it will not decay again. Because we have focused on $V_k = 20-100$~km/s, the relevant mass loss fractions are $f \approx V_k/c \sim 7\times 10^{-5} - 3\times10^{-4}$. Therefore, we maintain the masses of the particles in the simulations because the effect of the change in the kinetic energies of the particles due to the kicks is much greater than the change to the potential energy due to the decrease in particle mass. The light relativistic daughter particles escape from any system promptly after they are generated from the decay process and they are not of interest in this study. The light daughters have no effect on halo properties, and they have negligible effect on the expansion rate of the Universe and the growth of structure, even at late times, because their abundance is strongly suppressed by the small mass loss fraction~\citep{wang_zentner12}.

In order to make a direct comparison with prior simulations, we used the same initial conditions for both our uniform resolution simulations (B50) and our zoom simulations (Z12) as \citet{Rocha_etal12}. Moreover, we included a higher-resolution version of the same Galactic halo zoom-in simulation (Z13) with $\sim$ 1/8 times smaller particle mass for the highest resolution region in order to test convergence and to study the detailed internal structures of Galactic subhalos. All simulations have the same initial conditions as the fiducial CDM run starting at $z=250$. The cosmology used is based on WMAP7 results with $\Omega_M$=0.266, $\Omega_{\Lambda}$=0.734, $n_s$=0.963, $h$=0.71, and $\sigma_8$=0.801. In each case, we have identified halos using the publicly available Amiga Halo Finder (AHF)~\citep{Knollmann_etal09} code. The halo radius can be defined as the radius of a sphere
within which the average density is $\Delta_{\mathrm{vir}}$ times larger than the background density $\rho_{b}$ of the Universe:
\begin{equation}
\label{eq:halo_definition}  
\mathrm{M_{vir}} = 4 \pi/3 \rho_b \Delta_{\mathrm{vir}}(\mathrm{z}) r_{\mathrm{vir}}^3, 
\end{equation} 
where the $\Delta_{\mathrm{vir}}(\mathrm{z})$ depends on both the redshift and the given cosmology \citep{bryan1998}. The maximum circular velocity, $V_{\mathrm{max}}$, of a test particle within a halo is given by $V_{\mathrm{max}}\equiv {\mathrm{max}}\{[GM(< R)=R]^{1/2}\}$. The maximum circular velocity is achieved at a radius of $R_{\mathrm{max}}$. For an NFW profile, it is useful to note that the escape speed from the center of a halo is related to the maximum circular velocity by $V_{\mathrm{esc}} \approx 3 V_{\mathrm{max}}$. 
%We briefly describe the properties of each simulation here.  

The uniform resolution cosmological simulations (B50 series) track the evolution of $512^3$ particles in a cubic computational volume with a side length of $50\, h^{-1}\mathrm{Mpc}$. The dark matter particle mass is 9.83 $\times10^7 M_{\odot}$ with the force softening scale $\epsilon$=1.4 kpc. This set of simulations allows us to study the properties of group- or cluster-sized halos ($M_{\mathrm{vir}}\sim 10^{13}-10^{14}$ $M_{\odot}$), as well as overall large scale structure. We choose one set of decay parameters ($\Gamma^{-1}$ = 40 Gyr, $V_k$ = 100 km/s) to illustrate the effects of DDM on cosmological scales and on the structures of halos over a range of halo masses. As has been discussed previously in \citet{peter_etal10} and \citet{wang_zentner10}, and as we will demonstrate in~\S~\ref{section:results}, DDM significantly affects halos with circular velocities $V_{\mathrm{max}} \lesssim V_k$.

\begin{figure}
\includegraphics[height=6.0cm]{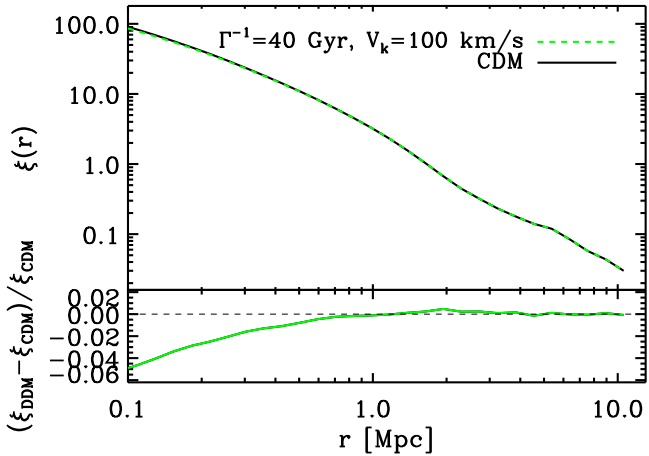}
\caption{
Dark matter two-point correlation function at $z=0$. The black solid line is from the CDM simulation, and the green dashed line is from the DDM simulation with $\Gamma^{-1}=40\, \mathrm{Gyr}$ and $V_k=100\, \mathrm{km/s}$. On large scales, the CDM and DDM simulations give nearly identical correlation functions with significant differences emerging on sufficiently small scales.
% This plot is the direct analogous to the left panel of Figure 3 in \citet{Rocha_etal12} for SIDM models.
}
\label{fig:corre_fn}
\end{figure}

%The halo selected for the zoom-in simulation series (Z12 \& Z13) was drawn from the 50$h^{-1}$ Mpc box. The computational power was then concentrated on the zoom-in regions.
In order to study such small halos, we select a halo from the 50 $h^{-1}$ Mpc box for a zoom-in simulation, wherein the computation power is focused on a single Milky Way-size host. For the CDM case, the selected host halo has $M_{\mathrm{vir}}$ = $10\time10^{12} M_{\odot}$ and $R_{\mathrm{vir}}$ = 265 kpc. For Z12, the smallest force softening scale was set to 143 pc in comoving units and the particle mass in the highest resolution region is 1.92$\times 10^5$ $M_{\odot}$. For Z13, the smallest force softening scale is 72 pc and the particle mass is 2.4$\times 10^4$ $M_{\odot}$. We note that our Z13 simulations have particle mass and force resolution comparable to Aquarius level 2 simulations \citep{Springel_etal08} and to Via Lactea I (VL-I) \citep{VLI}. Our Z12 simulations have a resolution between the level 3 and level 4 Aquarius simulations. Table 1 lists the properties of the simulations. For the zoom-in simulations the volumes listed refer to the number of virial radii in CDM run used to find the Lagrangian volumes associated with the zoom. The particle properties listed are for the highest resolution particles only. We run Z12 simulations for six different DDM models to sample from the parameter space that is not excluded by current Lyman-$\alpha$ forest data~\citep{Wang_etal13}. Due to limitation of computational resources we only run Z13 simulations for one DDM model ($\Gamma^{-1}$ = 10 Gyr, $V_k$ = 20 km/s) for convergence tests and demonstration purposes. We utilize the Z13 runs to test the completeness limit of subhalos self-consistently for our zoom-in simulations. The details of the comparison is shown in \S~\ref{subsubsection:convergence}. We find that for Z12 runs we resolve subhalos down to $V_{\mathrm{max}} > 8$ km/s with $M_{\mathrm{sub}} \gsim 2 \times 10^7 M_{\odot}$.
%and the same initial condition with Aquarius-A level 2 halo is adopted both in previous galactic subhalo studies for WDM models \citep{Lovell_etal12,Lovell_etal14} and SIDM models \citep{Vogelsberger_etal12}, which show to provide a reasonable resolution to resolve subhalo properties. 

We note that for the zoom-in simulations we only enable high resolution particles to decay. The more massive particles, which serve as boundary particles in the zoom-in simulations, behave as CDM particles. This approximation still reproduces the right gravitational environment for the zoomed galactic halos as we show in the next section using our uniform-resolution runs, so the effects of DDM models are confined well within Galactic scales.

\begin{figure}
\includegraphics[height=6.1cm]{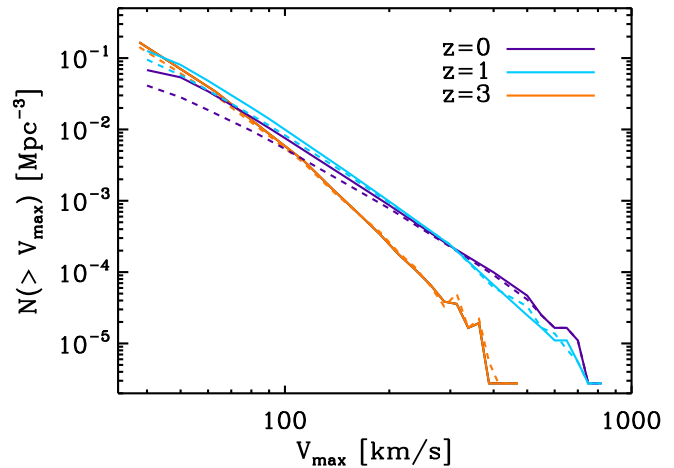}
\caption{
Cumulative number density of dark matter halos as a function of their maximum circular velocity ($V_{\mathrm{max}}$) at different redshift from B50-CDM (solid lines) and B50-t40-v100 (dash-dotted lines) cosmological simulations at $z=3$ (orange), $z=1$(blue), and $z=0$ (purple). The discrepancies between CDM and DDM do not show until later time, particularly for for $V_{\mathrm{max}} \lsim$200 km/s.
% This plot can be compared directly to the right panel of Figure 3. in \citet{Rocha_etal12} for SIDM models.% since we use the same initial conditions for the DDM simulations.
}
\label{fig:vfn_z}
\end{figure}
%---------------------------
\section{Simulation Results}
\label{section:results}
%---------------------------

We show a visual comparison of uniform-resolution CDM and DDM simulation snapshots in Figure~\ref{fig:snapshot}. The left panel is the line-of-sight projected density-squared drawn from the B50-CDM simulation, and the right panel is from the B50-t40-v100 DDM simulation with $\Gamma^{-1}$ = 40 Gyr, $V_k$ = 100 km/s. Visualized is a slice 10 $h^{-1}$Mpc deep and $50 \times 50 \, h^{-1}\mathrm{Mpc}$ on a side. The lack of obvious visual differences suggests that this DDM model agrees with CDM predictions on large scales and preserves the large-scale successes of the CDM model. 
%The plotting range is the size of the simulation box, 50 $\times 50 \, h^{-2}\mathrm{Mpc}^2$. The slice is a projection with a depth of 10 $h^{-1}$ Mpc along the line-of-sight direction. 

\begin{figure*}%
\includegraphics[height=8.4cm]{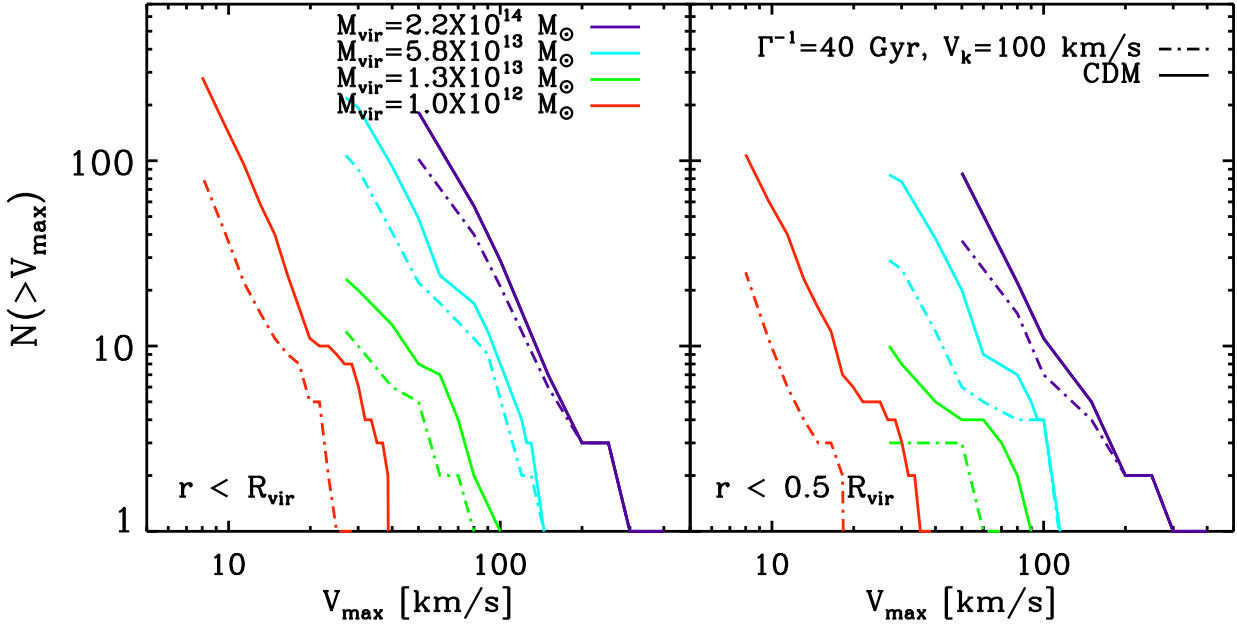}
\caption{ Subhalo cumulative velocity functions as a function of halo peak circular velocity ($V_{\mathrm{max}}$) for halos with mass range of $M_{\mathrm{vir}} = 2.2\times 10^{14} - 1.0\times 10^{12} M_{\odot}$ at {\it z}=0. Objects with $M_{\mathrm{vir}} \ge 1.3\times 10^{13} M_{\odot}$ are selected from our B50 simulation set, while the properties of halo with $M_{\mathrm{vir}} = 1.0\times 10^{12} M_{\odot}$ are derived from the Z12 zoom-in simulations. The solid lines represent the CDM results, while the dash-dotted lines are for the DDM model with $\Gamma^{-1}$=40 Gyr and $V_k$=100 km/s. This plot is directly analogous to the Figure 8. in \citet{Rocha_etal12} for SIDM models. The decay model efficiently reduces the number of halos with $V_{\mathrm{max}} \lsim$ 100 km/s. The velocity function difference between CDM and DDM models becomes more significant for smaller halos. For this particular model, the effect of decays is to uniformly reduce the number of halos with $V_{\mathrm{max}}$ $\ll$ 100 km/s, but the effect is smaller for larger $V_{\mathrm{max}}$.  The effect is stronger in the inner region for $r<$ 0.5 $R_{\mathrm{vir}}$ (right panel) than the overall effects in $r <$ $R_{\mathrm{vir}}$ (left panel).  
}
\label{fig:mass fn} 
\end{figure*} 

In Figure~\ref{fig:zoom_sim} we compare the Z12 CDM galactic halo simulations with two other DDM scenarios ($\Gamma^{-1}$ = 40 Gyr and $V_k$ = 100 km/s; $\Gamma^{-1}$ = 10 Gyr, $V_k$ = 20 km/s). Although the host galactic halos do not show obvious changes in density or shape, the effects on substructure are significant and depend upon the decay parameters. The substructure in each halo simulation exhibits similar broad spatial distributions, though the detailed and/or object-by-object comparisons are somewhat challenging because the potential structure of the host halo is altered by the DDM in each case and the character of satellite halo orbits depends sensitively on the potential structure \citep[e.g.][]{debattista_etal08,valluri_etal10,valluri_etal13}. The common characteristic of both DDM scenarios is that they show a reduction of subhalo abundance compared to CDM. For the case with $\Gamma^{-1}$ = 40 Gyr, $V_k$ = 100 km/s (middle panel), a significant number of small substructures remain, but their density profiles are shallower than the CDM case. It is evident that the densities achieved in the central regions of the subhalos in the DDM simulations are lower than those achieved in the CDM run. In the particular case of DDM with $\Gamma^{-1}$ = 10 Gyr, $V_k$ = 20 km/s (right panel), the smaller subhalos are almost completely washed out and there is a less significant change in the density profiles of large subhalos ($M \gsim$$10^8$$M_{\odot}$) relative to the case in which $\Gamma^{-1}$ = 40 Gyr, $V_k$ = 100 km/s.  

As we proceed to interpret our results, it will be useful to recall that the recoil speed $V_k$ determines the sizes of halos that will be significantly altered by the DM decays, with halos that have $V_{\mathrm{esc}} \lsim V_k$ most strongly affected by the decays and halos with $V_{\mathrm{esc}} \gg V_k$ minimally altered by the decays. Roughly speaking, a subhalo with a mass of $M \sim 10^8\, \mathrm{M}_{\odot}$ has $V_{\mathrm{max}} \approx 10\, \mathrm{km/s}$ and the escape speed from the center of the subhalo of $V_{\mathrm{esc}} \approx 30\, \mathrm{km/s}$. It is not surprising that a recoil speed of $V_k = 20\, \mathrm{km/s}$ significantly alters, or even completely unbinds, smaller halos. 

%---------------------------------------------------------
\subsection{Large-Scale Structure}
\label{subsection:LSS}

We begin the quantitative discussion of our simulation results with a brief examination of the large-scale structure within the CDM and DDM models. In Figure~\ref{fig:corre_fn} we show the two-point correlation functions of dark matter within the 50 $h^{-1}$ Mpc, uniform-resolution cosmological simulations. The fractional differences in the correlation functions between the CDM and DDM models is shown in the lower panel of Figure~\ref{fig:corre_fn}. On large scales $r \gsim 500\, \mathrm{Mpc}$, the CDM and DDM simulations agree well, as suggested by the visual comparison in Figure~\ref{fig:snapshot}. On scales $r \gsim 500\, \mathrm{Mpc}$, the differences between the CDM and DDM simulation correlation functions are less than 1\%. These differences increase to about 5$\%$ at a scale of $r \sim$ 100~kpc, and continue to increase toward smaller scales. For a typical galactic halo with virial mass $M_{\mathrm{vir}}$ $\sim$ $10^{12}$ $M_{\odot}$, the virial radius $R_{\mathrm{vir}}$ is $\sim$ 250 kpc and the escape speed from the halo center is $V_{\mathrm{esc}} \approx 600\, \mathrm{km/s}$, so the effects for the DDM scenario we study here are only important on sub-galactic scales.

%{\bf [I think it would be much better to have the velocity function be a distinct figure. It is better to use multi-panel plots only for things that are closely related.]} {\bf [MW: I separated the plots.]}
In Figure~\ref{fig:vfn_z}, we plot the cumulative number density of dark matter halos as a function of their maximum circular velocities ($V_{\mathrm{max}}$) at three different redshifts. The solid lines show the CDM results, and the dashed lines are from the DDM simulation. These two models agree well at $z=3$ at all $V_{\mathrm{max}}$ even below $V_{\mathrm{max}} \lsim 40\, \mathrm{km/s}$ ($V_{\mathrm{esc}} \lsim 120 \, \mathrm{km/s}$). Differences between the cumulative velocity functions emerge as we proceed to lower redshift. At $z=0$, the differences become significant for $V_{\mathrm{max}}$ $\lsim$ 200 km/s. For example at $V_{\mathrm{max}}$ = 100 km/s the decrement is $\sim$ 29$\%$ at $z=0$ and for $z=3$ it is just $\sim 6.5\%$. 

The result above is consistent with our earlier statement that the halo sizes most affected by DDM are determined by $V_k$. At all values of $V_k$, the impact of DDM is most significant at later times because more of the DDM will have decayed (and provided the requisite recoil speed to the stable daughter particles) at later times. The fraction of DM particles that have decayed at different redshifts is 
\begin{equation}
\label{eq:decay}  
\mathrm{N_{decayed}}(z)/\mathrm{N_{DM}} = 1-\mathrm{exp}[-\mathrm{t(z)}\Gamma^{-1}], 
\end{equation} 
where $t(z)$ is the physical time at redshift $z$. For $\Gamma^{-1}$ = 40 Gyr, at $z=3$ only $\sim$ 5 $\%$ of the DM particles have decayed. At $z=1$ this fraction is $\sim$ 14$\%$ and at $z=0$ it has risen to about $\sim 1/3$. This is an interesting feature of the DDM model because it suppresses small-scale structure growth most strongly at low redshift and therefore DDM models evade constraints on early structure formation that severely limit WDM models.

\begin{figure*} 
\includegraphics[height=8.0cm]{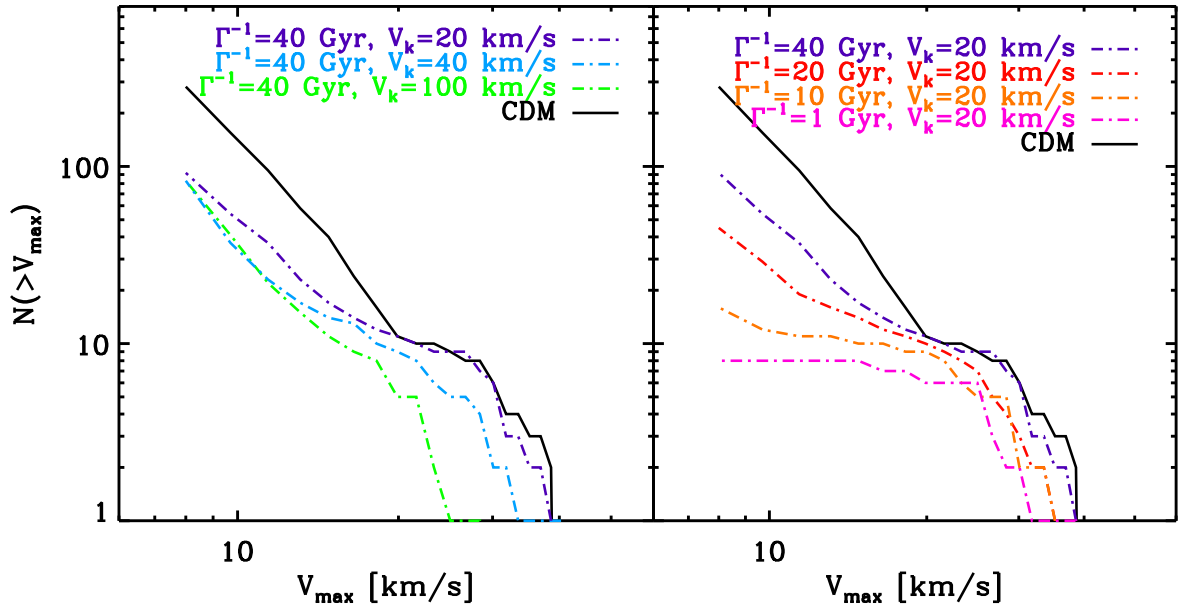}
\caption{ 
Subhalo cumulative velocity functions as a function of halo maximum circular velocity for galactic halo with $M_{\mathrm{vir}} = 1.0\times 10^{12} M_{\odot}$ for several different DDM models. These are derived from the Z12 zoom-in simulation sets. The solid black lines indicate the velocity function from the CDM simulation, and dash-dotted lines are from the DDM simulations. In the left panel, decay lifetime is fixed at $\Gamma^{-1}=40\, \mathrm{Gyr}$, and $V_k$ is allowed to vary between 20-100 km/s. In the right panel, $V_k$ is fixed at $V_k$ = 20 km/s, and $\Gamma^{-1}$ is allowed to vary between 1-40 $\mathrm{Gyr}$. 
}
\label{fig:vfn_Z12}
\end{figure*} 

We have used identical initial conditions to \citet{Rocha_etal12}, therefore our DDM results in Figure~\ref{fig:corre_fn} and Figure~\ref{fig:vfn_z} can be directly compared with Figure 3 in the SIDM simulations of~\citet{Rocha_etal12}. On large scales, both the DDM and SIDM two-point correlation functions show good agreement with the CDM model. However, for the dark matter density function, the SIDM model shows no significant differences {\em at all redshifts} while for the DDM model the discrepancies appear at $V_{\mathrm{max}} \lsim$ 200 km/s  and only {\em for z $\lsim$ }1.  \citet{Rocha_etal12} find that SIDM halos have constant-density cores across different mass ranges. However, these changes are restricted to the inner regions of the halos and $V_{\mathrm{max}}$ is not strongly affected within larger halos. % {\bf AP: do you know of any paper in which the WDM mass function as a function of redshift is explored?  It would be interesting to say something about that here, but for the life of me I can't find anything with a similar redshift range---the Schultz et al. paper out of the Irvine group only looks at z=6-13.} {\bf [I don't really know, but there is a paper using lensed high-redshift galaxies (z$\sim$10)from CLASH to constrain WDM. They used some analytical mass function though, not from simulation.]}

%{\bf [If you really want to study large-scale structure in an interesting way, you should also present the {\em halo} correlation functions in addition to the mass correlation functions. In other words, (1) make a few halo catalogs with cuts such as $V_{\mathrm{max}} \ge 400, 300, 200, 100\, \mathrm{km/s}$ (I just picked these numbers out of a hat, there may be more appropriate values given the size of the box and the resolution of the simulation). (2) Select all halos (including subhalos) above each cut to construct four catalogs from each simulation. (3) Plot the halo correlation functions in each of these cases compared to the DDM simulations. This would be a very interesting comparison and hopefully only shows significant deviations for the lowest $V_{\mathrm{max}}$ catalog. These are also more relevant to galaxy survey data.]} {\bf [MW: I think this might not be relevant to the focus of this paper, which is actually on galactic scale. The difference maybe too small to be observed by forthcoming galaxy surveys.]}

%-----------------------------------------------------------------------
\subsection{Halo Structure and Substructure}
\label{section:halo structure}

\subsubsection{Subhalo Abundace} 
\label{subsubsection:halo_abundace}

The previous section showed that the effects of our DDM models are confined to scales less than a few hundred kpc. In this section, we begin by examining the abundance of subhalos at $z$ = 0.  Figure~\ref{fig:mass fn} plots the subhalo cumulative velocity function as a function of $V_{\mathrm{max}}$ in a variety of host masses ($M_{\mathrm{vir}}$ = 2.2$\times$$10^{12}$ - 1.0$\times$$10^{14}$ $M_{\odot}$). The halos with $M_{\mathrm{vir}}$ $\ge$ 1.3$\times$$10^{13}$ $M_{\odot}$ are selected from our B50-CDM (solid) and B50-DDM-t40-v100 (dash-dotted, $\Gamma^{-1}$=40 Gyr and $V_k$=100 km/s) simulations. The halos with $M_{\mathrm{vir}}$ $=$ 1.0$\times$$10^{12}$ $M_{\odot}$ in these two panels are from the Z12 simulations.
%we focus on the small scale properties of halos in order to elucidate the effects of DDM. In Figure~\ref{fig:mass fn}  we show the subhalo cumulative velocity function as a function of $V_{\mathrm{max}}$ for halos across different mass ranges 

Figure~\ref{fig:mass fn} demonstrates that DDM scenarios induce different effects on subhalo abundance that become more significant for smaller halos. This behavior is different from the effects of SIDM models, which show only limited and mass-independent changes in subhalo abundance~\citep{Rocha_etal12}. The effects of DDM models start to become remarkable when $V_{\mathrm{max}}$ $\lsim$ 100 km/s, close to $V_k=100\, \mathrm{km/s}$ for this parameter set. The effects are also more dramatic in the inner regions of halos, as indicated in the right panel of Figure~\ref{fig:mass fn} for subhalos at r $<$ 0.5 $R_{\mathrm{vir}}$, than they are over the scale of the entire host halo (middle panel in Figure~\ref{fig:mass fn}).  As we will show, halos strongly affected by decay are less dense than their CDM counterparts, and are thus much more fragile to tidal disruption if they plunge deeply into a larger halo.

\begin{figure*}
\includegraphics[height=6.3cm]{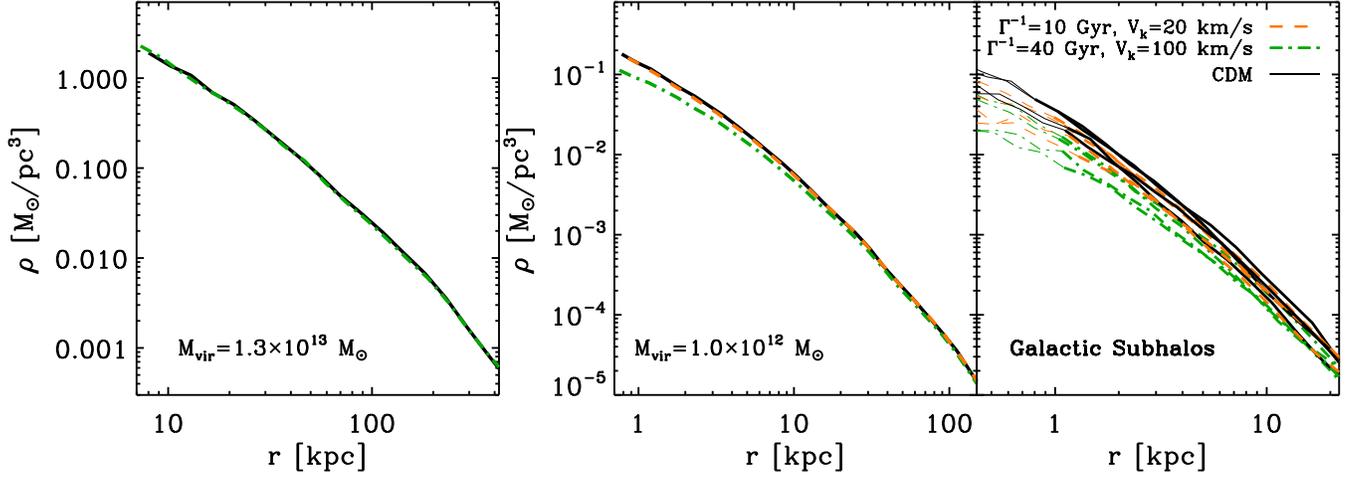}
\caption{Density profiles of halos of different masses at $z=0$. The left panel shows a $1.3\times10^{13}M_{\odot}$ object from the 50 $h^{-1}$ Mpc box cosmological simulations; the middle panel shows the galactic host halo in the Z12 simulations; the right panel shows the density profiles of the four largest subhalos, ranked by $V_{\mathrm{max}}$, of the Galaxy-sized host in the middle panel. The solid black lines indicate the density profiles from the CDM simulation; the orange dash-dotted lines are from DDM model with $\Gamma^{-1}$ = 10 Gyr, $V_k$ = 20 km/s; the green dash-triple-dotted lines are from the DDM simulation with $\Gamma^{-1}$ = 40 Gyr, $V_k$ = 100 km/s. For these parameters, the impact of the decay process is not significant on the density profile of a group-size object, but the effects become visible for halos below the Galactic scale. 
}
\label{fig:den}
\end{figure*}

Of course, the specific results of Figure~\ref{fig:mass fn} are relevant only for the DDM parameters $\Gamma^{-1}=40\, \mathrm{Gyr}$ and $V_k=100\, \mathrm{km/s}$ and varying both of these parameters can alter the influence of the DM decays. The results of Figure~\ref{fig:mass fn} are an illustrative example in which the effects of DDM are non-negligible across a wide range of subhalo sizes. In the following sections, we will focus on the effects of DDM on Milky Way substructure and explore a wider variety of DDM parameters. Indeed, the choice of DDM parameters $\Gamma^{-1}$=40 Gyr and $V_k$=100 km/s is likely an the extreme case. In a galactic halo, the number of subhalos is across-the-board reduced.  We will show that for models with $V_k \ll 100\, \mathrm{km/s}$, there will be interesting, but more subtle, scale-dependent effects on galactic halo substructure. 

The scale-dependence of the effects of decay on halo structure and substructure is related to the typical DM particle velocities in the halo with respect to $V_k$. \cite{peter_10} showed how the effects of DM decays depend on halo properties, namely the typical dynamical time ($t_{\mathrm{dyn}}$) and the typical speed of dark matter particles in the halo. ~\cite{peter_10} chose the viral speed $\left( V_{\mathrm{vir}} = \sqrt{{GM_{\mathrm{vir}}\over R_{\mathrm{vir}}}}\right)$ as the typical speed, and here we choose $V_{\mathrm{max}}$, which is typically close to $V_{\mathrm{vir}}$\footnote{$V_{\mathrm{max}} \approx 1.3 V_{\mathrm{vir}}$ for a halo with a concentration of $c=R_{\mathrm{vir}}/r_{\mathrm{s}}=12$ typical of MW sized halos}. For DDM with $V_k$ much greater than the typical speed of the DM halos ($V_k > V_{\mathrm{esc}} \approx 3 V_{\mathrm{max}}$), daughter DM particles will be ejected from the system due to their recoils and will cause the halos to lose mass. If $V_k \sim V_{\mathrm{max}}$, mass may be redistributed within halos making halo density profiles significantly less steep, but the fraction of mass ejected from the halos will be small. On the contrary, if $V_k \ll V_{\mathrm{max}}$ the DM decays will have minimal effect on halo mass and/or structure.  For the example of $\Gamma^{-1}=40\, \mathrm{Gyr}$ and $V_k=100\, \mathrm{km/s}$, $V_k \gg 3V_{\mathrm{max}}$ for all subhalos in the galaxy-sized halos in Figure \ref{fig:mass fn}, which is why the velocity function is uniformly suppressed.  For smaller $V_k$, we will expect to see a feature near $V_{\mathrm{max}} \sim V_k$ in the velocity function of substructure, owing to the difference in the effects of decay for $V_{\mathrm{max}} \lesssim V_k$ and $V_{\mathrm{max}} \gtrsim V_k$. 

As mentioned above, our goal is to focus on the DDM models that exhibit interesting phenomenology on galactic and sub-galactic scales, while not affecting larger-scale astrophysical limits. For galaxy groups and clusters (for $M_{\mathrm{vir}}\sim10^{13}-10^{15}M_{\odot}$ with $V_{\mathrm{max}}\sim$300-1000 km/s), all the DDM models in this study have $V_k$ $\ll$ $V_{max}$; for galactic halos ($M_{\mathrm{vir}}\sim10^{12}M_{\odot}$ with $V_{\mathrm{max}}\sim$150 km/s), only the model with $V_{k}$ = 100 km/s is comparable to the $V_{max}$ of the system. For the MW dwarf spheroidal satellite galaxies ($M_{\mathrm{sub}}\sim10^{8}-10^{9}$ $M_{\odot}$ with $V_{\mathrm{max}} \sim$ 10-20 km/s), the models with $V_{k}$ = 20 km/s have recoil speeds comparable to the range of the $V_{\mathrm{max}}$ values for the systems of interest.
 
\subsubsection{Milky Way Subhalo Abundace} 
\label{subsubsection:MW_halo_abundace}

In Figure~\ref{fig:vfn_Z12}, we focus attention on our high-resolution Galaxy-sized halos and show the subhalo velocity functions for several different DDM models. These curves show very different levels of suppression as a function of $V_{\mathrm{max}}$ for the various DDM parameters. In the left panel we consider the three models with $\Gamma^{-1}=40\, \mathrm{Gyr}$ and different values of $V_k$. From these results it is clear that the DDM models with larger $V_k$ more strongly suppress substructure with higher $V_{\mathrm{max}}$. In the limit of small $V_{\mathrm{max}} \lsim 10\, \mathrm{km/s}$, the kick velocity is significantly larger than $V_{\mathrm{max}}$ and all three simulations with $\Gamma^{-1}=40\, \mathrm{Gyr}$ yield similar subhalo abundances.  About 1/3 of these small subhalos remain. 

In the right panel, we compare all four models with $V_k=20\, \mathrm{km/s}$.  These models differ in $\Gamma^{-1}$.  This plot shows that DDM lifetime determines the level of substructure suppression at fixed $V_k$, and that suppression becomes dramatically more significant when $\Gamma^{-1} \lsim H_0^{-1}$. It is interesting to notice that for $V_{\mathrm{max}} \lsim$ 10 km/s, the subhalo number seems to be proportional to the decay lifetime and does not depend on the kick velocity. This is because the smallest recoil speeds in our simulations are still greater than the circular velocities of these systems, so the decay lifetime sets the fraction of particles ejected and therefore determines the suppression of small subhalos. In the limit that $V_k \gg V_{\mathrm{max}}$ the effect of DDM is determined almost entirely by the lifetime of the particles. For $\Gamma^{-1}$ = 10 Gyr, about 1/10 of the small subhalos remain, while for $\Gamma^{-1}$ = 1 Gyr only $\sim$ 10 subhalos with $V_{\mathrm{max}} \gsim$ 10 km/s remain within the system at $z=0$.

 As we discussed earlier, if the kick velocity is much greater than the typical circular velocity of the DM halo, DM particles are ejected from the system. As discussed in~\cite{peter_10}, if the decay lifetime is greater than dynamical time scale of the halos, the halos will go through adiabatic expansion to accommodate changes to the gravitational potential. This will result in a reduction of halo size and a shift in the velocity function. For our galactic zoom-in simulations, the DDM model with $\Gamma^{-1}$=40 Gyr and $V_k$=100 km/s matches this criterion for all the galactic subhalos in the system. We find that the behavior of the simulated subhalo velocity function agrees well with the analytical expectation. 
 
 However, when comparing to observations, it is likely that a uniform reduction of the Galactic subhalo velocity function at all circular velocities cannot describe the Milky Way satellites because a reasonable number of massive subhalos are still required to host the observed Milky Way satellites (i.e. \cite{Koposov_etal08}).  Figure \ref{fig:vfn_Z12} shows that models with decay times $\Gamma^{-1}\lesssim H_0^{-1}$ suppress the subhalo velocity function too severely even for a small $V_k \sim 20\hbox{ km/s}$.  In order to match our simulations with current observational data, it is more reasonable to consider DDM models with $V_k$ $\lsim$ 40 km/s and $\Gamma^{-1}$ $\lsim$ 40 Gyr. This agrees with the results in~\cite{peter_benson10}, who examine the effects of DDM on the number of Milky-Way satellites using semi-analytical models. They show that the most relevant parameter range is around $V_k \sim$ 20-200 km/s for $\Gamma^{-1} \lsim$ 30 Gyr.  We will defer a more rigorous study of \emph{constraints} on DDM based on Galactic substructure to future work.

%We note that it is unlikely to have spurious halos form in the DDM simulation in a similar way as those that form in the WDM or HDM case~\citep{Bode_etal01, Wang_etal07,Lovell_etal14}. This is true even in the case of a relatively short lifetime 
We note that the formation of spurious halos,  as observed in WDM and HDM~\citep{Bode_etal01, Wang_etal07,Lovell_etal14} is unlikely in DDM simulations, even for relatively short lifetimes ($\Gamma^{-1}$ $\lsim$ a few Gyr). In the case of a decay lifetime shorter than the age of the Universe, small-scale structure below the free-streaming scale will be washed out by $z=0$ and power spectra will trend toward pure WDM models. However, the time evolution of DDM structure growth is very different from thermal-relic WDM, thus the evolution of small halos is very different from what is found in the WDM literature (i.e. \cite{Angulo_etal13}). Small halos in DDM models still assemble much of their mass early in their evolution, just as in CDM, but they are continuously perturbed by the decay process. At the stage when all the DM particles have decayed, the small halos below the free-streaming scale are completely destroyed. In contrast, in WDM, the spurious small halos are formed from the fragmentation of filaments originating from the initial, unperturbed particles~\citep{Wang_etal07}. Thus our DDM simulations cannot be affected by this numerical issue that plagues WDM and HDM models.

%------------------------------------------------------------------
\subsubsection{Halo Density Profiles}
\label{subsubsection:density_profiles}

In Figure~\ref{fig:den} we show the density profile of halos as a function of mass and of decay parameters at $z=0$. In the left panel we show a galaxy-group-size halo with $M_{\mathrm{vir}} = 1.3\times10^{13}$ $M_{\odot}$ selected from the B50 simulations. In the middle panel we show profiles of the host halos in the Z12 simulations with $M_{\mathrm{vir}} = 1.0\times10^{12}$ $M_{\odot}$, while in the right panel, we display the density profiles of the four largest subhalos (by $V_{\mathrm{max}}$) of the $1.0\times10^{12} M_{\odot}$ host, again in the Z12 simulations. The solid black lines indicate the density profiles from the CDM simulation, the orange dashed lines are from DDM model with $\Gamma^{-1}$ = 10 Gyr, $V_k$ = 20 km/s, and the green dash-dotted lines are with $\Gamma^{-1}$ = 40 Gyr, $V_k$ = 100 km/s.  These two decay models bracket extremes for the effects on the subhalo velocity function (Figure \ref{fig:vfn_Z12}).  For the Galactic subhalos, the thick lines are plotted down to the convergence radius derived using the methods in~\citet{Power_etal03} and the thin lines are plotted down to three times the softening scale. For the galactic host halo and galaxy group (the left two panels), the density curves are plotted down to convergence radius in~\citet{Power_etal03}. 
%profiles of the four subhalos of the Galactic halos in the middle panel that have the largest $V_{\mathrm{max}}$. 

%--------------------------------------------------------------------
 \begin{figure}
\includegraphics[height=8.0cm]{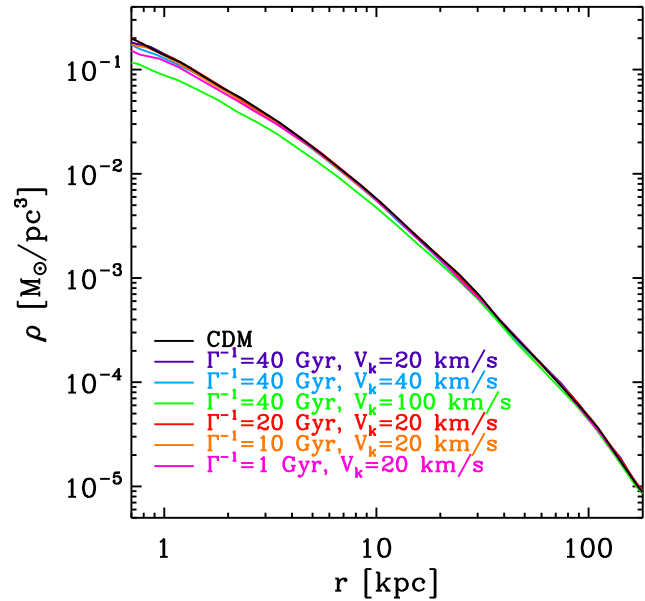}
\caption{Density profile of the host halos in the Z12 simulations at $z=0$. For most of the decay models we consider here show little density suppression concentrated on the central region, except the case with $\Gamma^{-1}$ = 40 Gyr, $V_k$ = 100 km/s.  This model has a relative large $V_k$ relative to other models, and is comparable to the virial speed of the host halo.
}
\label{fig:den_host_tot}
\end{figure}

It is clear that while these decay models have negligible impact on the structure of groups and clusters, the effects become more significant in smaller objects. This reflects the fact that the value of $V_k$ should be close to or greater than the typical $V_{\mathrm{max}}$ of the halos to produce significant changes. For the galactic main halo density profiles in the middle panel in Figure~\ref{fig:den}, we can see that the model with small kick velocity ($V_k$ =20 km/s) has very small effects on the host halo density profiles. For example at $r$=2 kpc the density suppression is about 5$\%$. Even in the case of $\Gamma^{-1}$ = 10 Gyr, where about $70\%$ of the DM has decayed and received a concomitant kick, the effect is still negligible. If we increase the kick velocity, for example up to 100 km/s, those particles which have decayed will receive a kick velocity comparable to the dynamical scale of the system ($V_{\mathrm{max}}$ =158 km/s), and the density profile can have obvious deviations from the CDM case even if only 30 $\%$ of the DM particles have decayed (as with $\Gamma^{-1}$ = 40 Gyr). For example at $r$=2 kpc the density deviation in the $V_k=100\, \mathrm{km/s}$ model is roughly 29$\%$, which is about six times of the suppression from the $V_k$ =20 km/s case. 

In Figure~\ref{fig:den_host_tot} we plot the galactic main halo density profiles for all six DDM models that we use to simulate the Z12 halo. Most of the DDM models we explore exhibit only limited deviations from CDM and what deviations do exist are generally restricted to the central regions of the halo. This reflects our choice of parameters, which in turn is driven by constraints on DDM models from large-scale structure \citep{Wang_etal13}. For the model with the largest kick speed, $V_k=100$~km/s, particles may be ejected from the halo or placed on significantly more extended, highly eccentric orbits. This alteration of dark matter particle orbits results in a significant change to the host halo density profile.    
    
In contrast to the main halo, the right panel of Figure~\ref{fig:den} indicates that DDM at relatively small kick velocities, not in tension with existing constraints \citep{Wang_etal13}, can significantly alter Galactic subhalo density profiles (in addition to their abundance). The objects depicted in the right panel of Figure~\ref{fig:den} have masses of $M_\mathrm{{sub}} \sim$ $10^{9}$ - $10^{10}\, \mathrm{M}_{\odot}$, in the CDM case, which corresponds to $V_{\mathrm{max}}$ $\sim$ 30-40 km/s. For the DDM model with $V_k=100\, \mathrm{km/s}$ ($V_k > V_{\mathrm{max}}$ for Galactic subhalos), those DM particles that receive the kick after decay will either have very eccentric orbits or become unbound from the subhalos. Consequently, in this scenario the effects of DDM on density profiles is not restricted to the central regions of the subhalos. We will return to subhalo density profiles in \S~\ref{subsubsection:convergence}, where we will show that simulation resolution is important for the robustness of the subhalo properties in DDM simulations, and address possible resolution affects on the convergence of subhalo density profiles.
 
\begin{figure*}
\includegraphics[height=8.3cm]{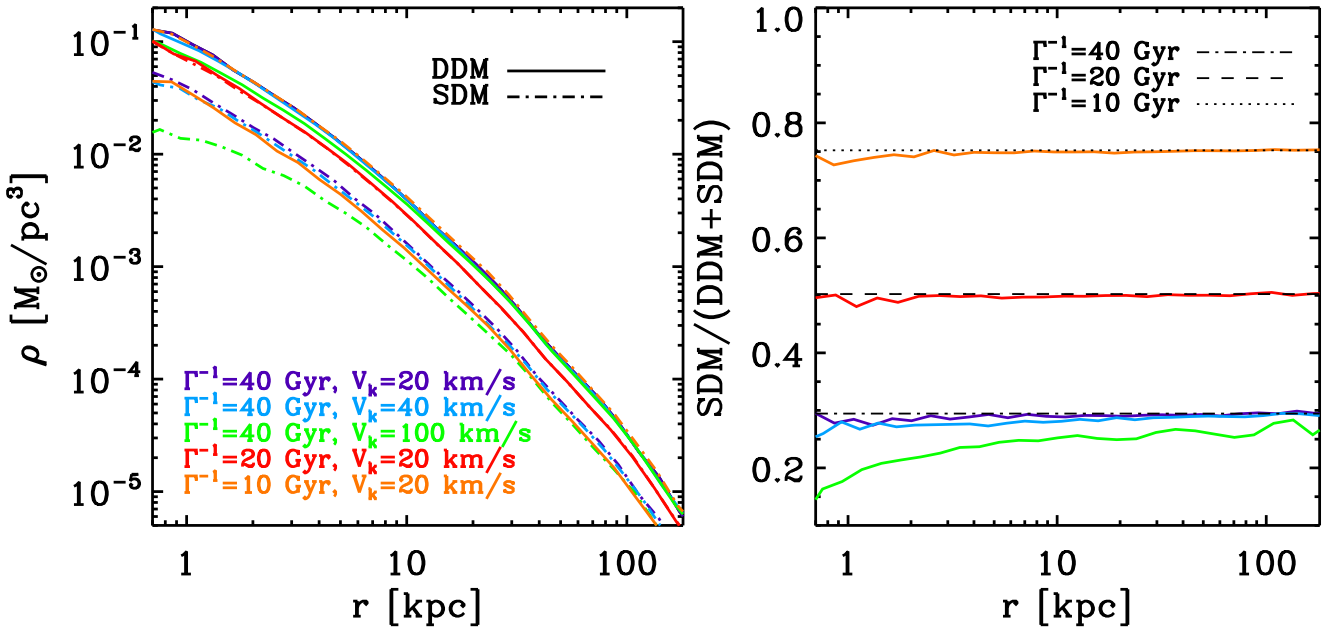}
\caption{ \textit{Left panel}: Density profiles of the host halos in the DDM simulations. We show the DDM components (solid lines, for DM particles before decay) and SDM components (dash-dotted lines, for DM particles after decay) separately for the Z12 simulation series at $z=0$. For $\Gamma^{-1}$ = 1 Gyr, essentially all of the DDM particles have decayed by $z=0$, so we do not show these profiles here. Similarly, note that the DDM profiles have a higher amplitude for each case with $\Gamma^{-1}=40\, \mathrm{Gyr}$, but that the reverse is true for the cases with shorter lifetimes. \textit{Right panel}: The fractional contribution of SDM component to the total density as a function of radius for different DDM models. The color scheme is the same as the left panel. The predicted SDM fraction using Eq.~\ref{eq:decay} and ignoring the amplitude of the kick speed is plotted with a black dotted line for $\Gamma^{-1}$ = 10 Gyr, a dashed line for $\Gamma^{-1}$ = 20 Gyr, and a dash-dotted line for $\Gamma^{-1}$ = 40 Gyr. 
}
\label{fig:den_host}
\end{figure*}
 
In Figure~\ref{fig:den_host}, we show the Galactic host density profiles for the different DM components in the Z12 simulations. The solid lines show the contributions from the ``cold," component (DDM), which is made up of the parent DM particles that have not yet decayed. The dash-dotted lines show the ``warm" component (SDM), which is comprised of the daughter DM particles that have received recoil kick velocities. We display DDM models with three different decay lifetime values: $\Gamma^{-1}$ = 40 Gyr, for which $\sim$ 30 $\%$ of the DM particles have decayed at $z=0$; $\Gamma^{-1}$ = 20 Gyr, for which $50\, \%$ of the DDM have decayed; and $\Gamma^{-1}$ = 10 Gyr, for which $\sim$70 $\%$ of the parent, DDM particles have decayed by $z=0$. In the $\Gamma^{-1}$ = 1 Gyr model, nearly all the DDM particles have decayed by $z=0$, so we do not show this model in Figure~\ref{fig:den_host}. In the right panel of Figure~\ref{fig:den_host}, we plot the fractional contribution of SDM to the halo density profiles and also the analytical prediction by Eq.~(\ref{eq:decay}), which ignores the dynamical effects introduced by the kick speed, at $z=0$. 

From the left panel of Figure~\ref{fig:den_host}, it is evident that the decay lifetime essentially determines the normalization of the density profiles for the DM components. This is also shown in the right panel of Figure~\ref{fig:den_host} for which the fraction of DDM is largely uniform across the entire halo for most of the models. While the distribution of DDM and SDM generally follows closely the original, unaltered halo profile of the CDM simulation, the distribution of SDM and DDM particles can be affected significantly for large values of the kick speed, $V_k$. For $V_k=100\,\mathrm{km/s}$, the SDM takes on a significantly less centrally-concentrated density profile than its CDM or the other DDM counterparts. The parent DDM particles respond to this reduction in the gravitational potential and also exhibit a significantly shallow density profile than the DM in the CDM case or the DDM cases with lower kick speeds. The right panel of Figure~\ref{fig:den_host} shows that the fraction of SDM is reduced by $5\%$ on all radii and goes down to 15$\%$ at the central region. For $V_k=40\, \mathrm{km/s}$, the contribution of SDM also has mild decrement of 2-5$\%$ for $r\lsim$ 10 kpc. As discussed in~\citet{Sanchez-Salcedo_03}, {\em on average} the net effect of decays is to impart an amount of energy $\Delta E \approx m V^2_k/2$ on the dark matter, independent of the initial velocity. The changes in average kinetic energy result in changes in particle orbits, causing an expansion of dark matter halos and making dark matter profiles more shallow. Previous studies find that this analytical model reproduces the properties of isolated simulations well~\citep{peter_etal10,wang_zentner12}.

We note the similarity of our galactic density profile feature with CWDM simulations \citep{Anderhalden_etal12}, which also contain a mixture of cold and warm components. However, as pointed out in \cite{peter_etal10}, DDM makes distinct predictions relative to CWDM because the density profiles of DDM models exhibit strong time evolution. This is expected because the ratio of the two DM components is controlled by the decay lifetime, and so the effects on the overall halo density profiles become more profound at later times. 

\begin{figure*}
\includegraphics[height=8.3cm]{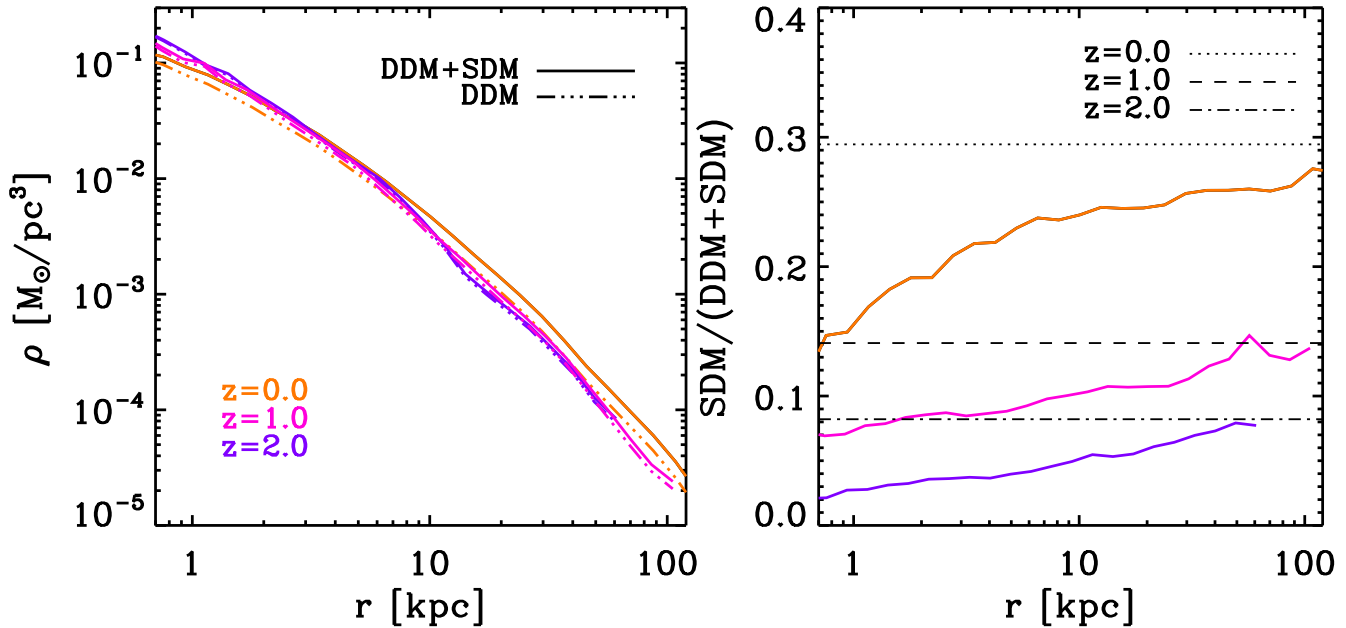}
\caption{ \textit{Left panel}: Density profile of the host halos for the all DM component (solid lines) and DDM component (dash-triple-dotted lines, for DM particles before decay) in the Z12 simulations with $\Gamma^{-1}$ = 40 Gyr, $V_k$ = 100 km/s at $z=0$ (orange), 1.0 (pink), and $2.0$ (purple). \textit{Right panel}: The fractional contribution of DDM component to the total density as a function of radius at different redshift. The color scheme is the same as the left panel. The predicted DDM fraction is plotted with a black dotted line for $z=0$, a dashed line for $z=1.0$, and a dash-dotted line for $z=2.0$.
}
\label{fig:den_host_z}
\end{figure*}

We examine the redshift dependence of the density profile in Figure~\ref{fig:den_host_z}, which plots total halo density profiles as well as the relative contributions from the DDM and SDM components for the model with $\Gamma^{-1}$ = 40 Gyr, $V_k$ = 100 km/s. In the right panel, we can see the strong redshift evolution of the SDM contribution. More interestingly, the radial distribution of the SDM fraction deviates significantly from the prediction from Eq.~\ref{eq:decay} at each redshift, which are shown by black lines in different line styles, on the central region, and tends to agree better at the outer region. This redshift-dependent behavior exists in all the DDM models unless the decay lifetime is very short. At the same time, the scale-dependent behavior exists for DDM models in which $V_k$ is comparable to the dynamical speeds within the system of interest.
%In Figure~\ref{fig:den_host_z}, we show total halo density profiles as well as the relative contributions from the DDM and SDM components for the model with $\Gamma^{-1}$ = 40 Gyr, $V_k$ = 100 km/s. 

%---------------------------------------------------------------------------------
\subsection{Milky Way Substructure}
\label{subsection:subhalos}

%-----------------------------------------------------

\subsubsection{Convergence Tests}
\label{subsubsection:convergence}

\begin{figure}%[ht]
\includegraphics[height=7.3cm]{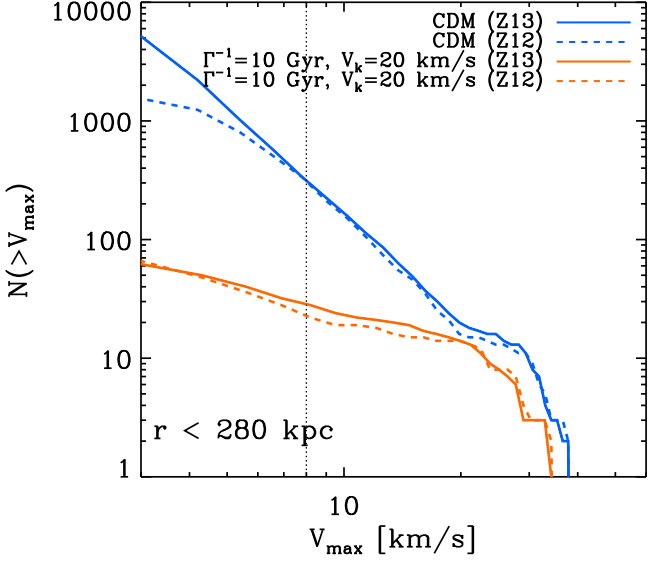}
\caption{ 
Resolution test of subhalo cumulative velocity function for the galactic zoom-in simulations (Z12 v.s. Z13). Here we show the cumulative subhalo abundance as a function of $V_\mathrm{{max}}$ within 280 kpc for both the CDM and DDM models ($\Gamma^{-1}$ = 10 Gyr, $V_k$ = 20 km/s) with two different resolutions for self-consistent comparison. The blue lines show the CDM case, and the orange lines represent the DDM case. For each model the high-resolution one (Z13) is shown with solid lines and the low-resolution on (Z12) is in dash lines. The horizontal dotted line indicates where the CDM velocity functions start to disagree, which corresponds to $V_\mathrm{{max}} \sim$ 8 km/s.  
}
\label{fig:vfn_res} 
\end{figure}

In order to study the demographics and internal structures of subhalos in a Milky Way-like halo, we focus on the results from our Z12 and Z13 zoom-in simulations in this section. To address convergence, we simulated one galactic halo with two different resolutions (the Z12 and Z13 simulations). We begin this section with a convergence test on the subhalo cumulative velocity function to examine the completeness of the Z12 zoom-in simulations, which is the resolution we adopted to demonstrate the effects of DDM models on galactic subhalo abundance in Figure~\ref{subsubsection:MW_halo_abundace}. Figure~\ref{fig:vfn_res} shows the results of this comparison between the Z12 and Z13 simulations within 280 kpc of the host galactic halo. We use a slightly more extended radius than the virial radius ($\sim$ 265 kpc) to account for the mild subhalo position change due to phase offsets between the resolutions. The CDM runs are shown as in blue curves while the DDM runs with $\Gamma^{-1}$ = 10 Gyr, $V_k$ = 20 km/s are shown in orange. The solid lines indicates the higher-resolution runs (Z13) for each DM models, and the lower resolution runs (Z12) are shown with dash lines. Resolution effects reduce the number of small subhalos in the Z12 runs for the CDM case, and the agreement extends down to $V_\mathrm{{max}} \gsim$ 8 km/s \citep{ELVIS}. Because DDM models usually efficiently eliminate small substructures, it is subject less to the resolution limits in terms of cumulative halo abundance and shows good agreement down to $V_\mathrm{{max}} \gsim$ 8 km/s. However, to be conservative we quote the results from the CDM runs and apply the completeness limits of $V_\mathrm{{max}} \gsim$ 8 km/s in all our Z12 subhalo velocity function studies in \S~\ref{subsubsection:MW_halo_abundace}.

%The we will briefly discuss the numerical convergence of the internal properties of the Galactic subhalos. 
In Figure~\ref{fig:subhalo_den} we illustrate the dependance of subhalo density profiles on the numerical resolution. The top three panels of Figure~\ref{fig:subhalo_den} show the three subhalo density profiles from two CDM simulations with different resolutions (Z12-CDM $\&$ Z13-CDM), while the lower three panels are their DDM counterparts with $\Gamma^{-1}$ = 10 Gyr, $V_k$ = 20 km/s. Those density profiles are plotted up to the convergence radius suggested in~\citet{Power_etal03} using thick lines. We extend these profiles down to a halo-centric radius of 2 times the softening scale with thin lines. The vertical dotted lines show 2.8 times the softening scale, outside of which the gravitational force is Newtonian. The CDM simulations at different resolutions agree well all the way down to a radius of 2.8 times the softening scale and clearly match well for all radii greater than the convergence radius of \citet{Power_etal03}. On the other hand, the lower-resolution (Z12-t10-v20) and higher-resolution (Z13-t10-v20) DDM simulations start to deviate from each other even for scales close to, or slightly larger than, the Power convergence radius. This shows that resolution is particularly important to resolve Galactic subhalos in DDM scenarios. 

The reasons why DDM subhalo structure converge more slowly than CDM case can be understood as follows. Each DM particle in the simulation actually represents a group of DM particles. In a small volume, where the number of DDM particles within the simulations is relatively small, the ratio of parent and daughter particles may not reflect the global value. Furthermore, the number of kick velocity directions sampled through the decays can be limited if sampled only by a small number of particles. These errors are sampling errors and scale inversely with the square root of the particle number. Consequently, density profiles are expected to be noisier in the interiors of DDM halos than of their CDM counterparts because there are fewer particles in these regions and these particles may sample the decay process poorly. For example, in Galactic subhalos with $M_{\mathrm{sub}}$ $\gsim$ 5$\times10^{8}$ $M_{\odot}$ in the Z12-CDM run, there are 200-700 DM particles for $r \lsim$ 0.7 kpc, while for Z13 the particle numbers are $\sim 3000-5000$. We will restrict our analysis related to subhalo inner structure to the Z13 runs to make sure the results are robust.
%The second reason that DDM subhalo structure may be more poorly resolved than CDM subhalo structure is that in the inner regions of DDM halos the densities tend to be lower due to the decays and recoils. The expected two-body collision relaxation time in the DDM scenario will be smaller ($t_{\mathrm{relax}} \propto N/\ln N$) than in the standard CDM scenario, and the inner regions of halos will be more susceptible to relaxation effects. Consequently, DDM halos require more particles to fully resolve the inner regions of halos. This effect is expected to be more severe for smaller objects or objects with shallower potentials. 

\subsubsection{The Density Profile Shape of Subhalos}
\label{subsubsection:density_subhalos}

\begin{figure*}%[ht]
\includegraphics[height=10.0cm]{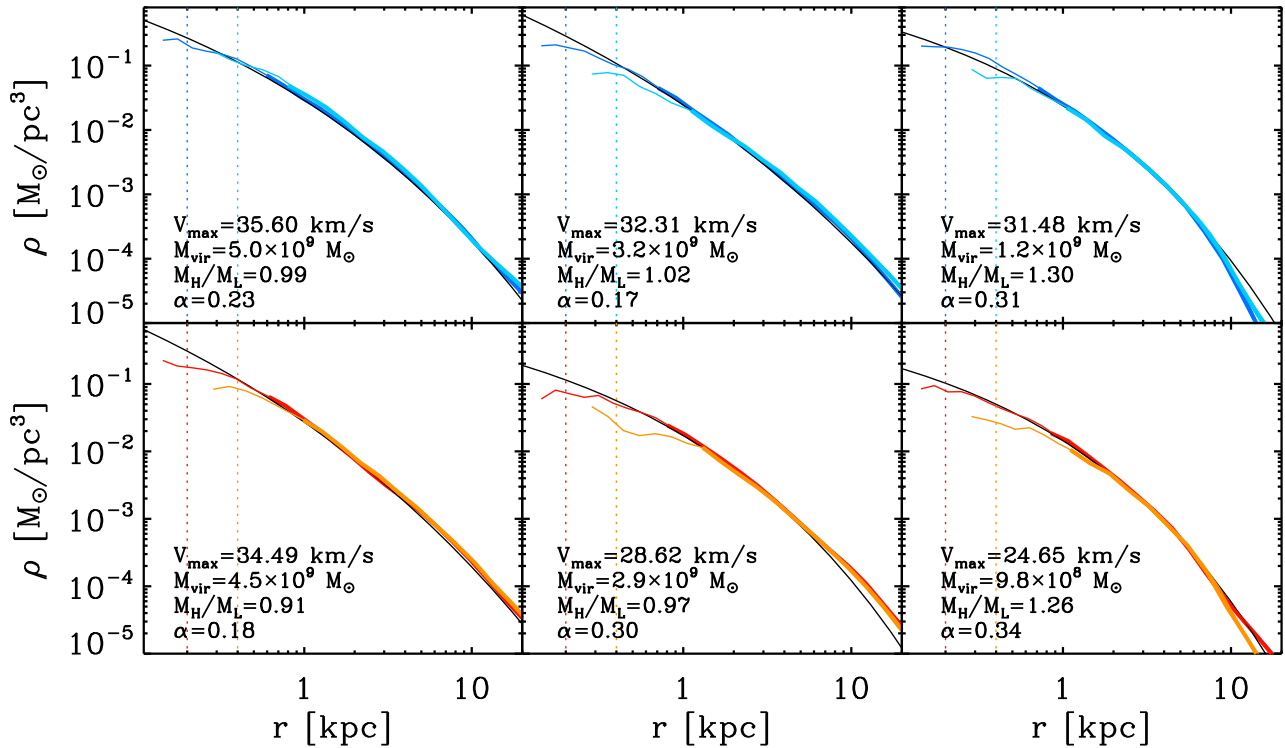}
\caption{ 
Density profiles for subhalos matched between the high (Z13) and low (Z12) resolution for the simulations with CDM and DDM model with $\Gamma^{-1}$ = 10 Gyr, $V_k$ = 20 km/s. The top three panels show the CDM simulation results, and the lower three panels are from the DDM simulations. In the CDM case the light blue corresponds to low and navy blue corresponds to high resolution, while in DDM case the orange corresponds to low and red to high resolution. "$M_H/M_L$" represents the ratio of the subhalo mass between the high and low resolution density profiles. They are plotted up to the convergence radius from \citet{Power_etal03} with thick lines, and extended to 2 times of the softening scale ($\epsilon$)with thin lines. The vertical dotted lines show 2.8 $\epsilon$, where the gravitational force becomes Newtonian. The solid black lines show the best-fit Einasto profile for the Z13 simulations. The value of $\alpha$ for the Einasto fit is also shown in each panel.
}
\label{fig:subhalo_den} 
\end{figure*}

Previous studies show that the inner density profiles of subhalos are well described by the Einasto form \citep{Springel_etal08},  
\begin{equation}
\label{Einasto}  
\rho(r) = \rho_{-2} \mathrm{exp} \Big( - {2 \over \alpha} \Big[\Big({r\over r_{-2}}\Big)^{\alpha}-1\Big]\Big),
\end{equation} 
where $\rho_{-2}$ and $r_{-2}$ are the density and radius at the point where the local slope $\mathrm{d}\ln \rho/\mathrm{d}\ln r = -2$. We test to determine if our subhalos have this similar form and plot the density profiles in Figure~\ref{fig:subhalo_den}. This figure indicates that our subhalo profile shapes are changed due to the DM decay process. From left to right, which corresponds to moving from more massive to less massive subhalos, it is apparent that the reduction of the inner density due to the decay process becomes more significant. We attempt to capture this feature by fitting the profiles from Z13 simulations with Einasto profiles and observe how the value of $\alpha$, which indicates how fast the slope of profile changes, varies with different DM models. 

We fit our Z13 subhalo density profiles with Einasto profiles over the radial range of 800 pc up to an outer radius. The value of outer radius is the smaller of 5 kpc or 1.5 $R_{\mathrm{max}}$. The choice of the lower limit, 800 pc, is comparable to the average Power convergence radius \citep{Power_etal03} of the 15 most massive subhalos (ranked by $V_\mathrm{{max}}$) in the Z13 host halo. We derive the fit by minimizing the goodness-of-fit measure:
\begin{equation}
\label{goodness_of_fit} 
Q^2={1\over N_{bins}} \sum^{N_{bins}}_{i=1} [\mathrm{log } \rho_{\mathrm{subs}}(r_i)- \mathrm{log } \rho_{\mathrm{model}}(r_i)]^2
\end{equation}
where $\rho_{\mathrm{model}}$ here represents the Einasto profile.

Below when we examine the subhalo circular velocity curves in ~\S~\ref{subsubsection:dSph}, we will adopt a procedure similar to the one described in \citet{Boylan-Kolchin_etal12} to correct for the force softening effects. We apply the best-fit Einasto profile derived using the method describe above to model the matter distribution for $r <$ 800 pc and use the simulated results for $r >$ 800 pc. We note that we apply the correction to a more extended range ( $r <$ 800 pc) than the one in \citet{Boylan-Kolchin_etal12} ($r <$ 291 pc) so that the extrapolation is based on the fitting results that are robust in the two-body relaxation criterion in \cite{Power_etal03}. %We explore a few different fitting lower bound between 291- 800 pc for the circular velocity profile correction, and we find that most of the subhalos are not sensitive to different choices. %The results are demonstrated in  ~\S~\ref{subsubsection:dSph}.
%Later we will apply this Einasto profile fit to the subhalo circular velocity curves for $r <$ 800 pc when we compare them with the Milky-Way dSph data in ~\S~\ref{subsubsection:dSph}. Although this method is similar to the one adopted in \citet{Boylan-Kolchin_etal12} to correct for the effects of force softening, we fit the profiles in a more extended range which has lower limit greater or comparable to the Power convergence radius of the subhalos we examine here.

Each panel in Figure~\ref{fig:subhalo_den} shows the best-fit Einasto profile for the Z13 simulations with a black solid curve. We also mark the value of best-fit $\alpha$ for each subhalo. We can see that Einasto profile fits provide an excellent description of the inner subhalo DM distribution. It is clear that in this DDM scenario, the reduction of DM density is more significant in the inner region than the outer region, and also the amount of suppression is more significant for smaller objects. This is also reflected on the value of best-fit $\alpha$, where on average DDM subhalos tend to render slightly larger $\alpha$ than their CDM counterparts.

%In the CDM cases, the values of $\alpha$ inferred from the fits are not very different from the NFW equivalent, which is $\alpha$ = 0.22. However, in the DDM simulations the inferred values of $\alpha$ increases markedly as subhalo mass decreases, and $\alpha=0.34$ for the smallest subhalo shown in the Figure, with $M_{\mathrm{sub}}$ $\sim$ 1.0$\times 10^9 M_{\odot}$). It is clear that in this DDM scenario, the reduction of DM density is more significant in the inner region than the outer region, and also the amount of suppression is more significant for smaller objects. 

The average central densities of dSphs are constrained by stellar kinematical data, which give $\rho \sim$ 0.1 $M_{\odot}/pc^3$ within a radius of $r=300$~pc ~\citep[e.g.,][]{Strigari_etal08}. Both our CDM and DDM simulations are broadly consistent with these constraints. For comparison, using the Einasto profile fits we find that the average central density for the largest subhalos in our Z13 DDM simulation at 0.3 kpc is about $\rho \approx 0.06 - 0.17\, \mathrm{M}_{\odot}/pc^3$ for $M_{\mathrm{sub}} \gsim 5\times10^{8} M_{\odot}$. For subhalos with $M_{\mathrm{sub}} \sim 1\times10^{8} - 5\times10^{8} M_{\odot}$, the central density is reduced to $\rho \approx$ 0.015 - 0.05 $\mathrm{M}_{\odot}/pc^3$. In $\Lambda$CDM simulations, subhalos which host the dSphs likely lie the mass range $\sim 2 \times 10^{8}-2 \times10^{9}\, \mathrm{M}_{\odot}$~\citep{Strigari_etal10, Boylan-Kolchin_etal12}. However it is important to note that these masses were inferred for $\Lambda$CDM halos only, and numerical simulations with different DM properties may predict somewhat different results because the structure of subhalos has changed. A more detailed analysis requires matching observed dSph stellar kinematic data to our set of DDM subhalos and we will pursue such a detailed analysis in a forthcoming follow-up paper. %{\bf need to change this number}

\subsubsection{$V_{\mathrm{max}}$ v.s. $R_{\mathrm{max}}$} 
\label{subsubsection:vmax_rmax} 

It has become common practice to characterize the sizes and internal structures of dark matter subhalos using the maximum circular velocity $V_{\mathrm{max}}$, and the radius at which this maximum circular velocity is achieved $R_{\mathrm{max}}$. This is at least in part because these quantities fully characterize an NFW profile and can be measured reliably in simulations because they are not sensitive to idiosyncratic choices in fitting procedures. In Figure~\ref{fig:vmax_rmax}, we show the $R_{\mathrm{max}}$ vs. $V_{\mathrm{max}}$ scatter plot for Galactic substructure from our Z13 simulations. We can see that the distribution of DDM subhalos is very different than the CDM case, which again indicates that the subhalo structure has been altered significantly in this DDM scenarios. The subhalos in our CDM simulations show good agreement with the previous study from the Aquarius simulations~\citep{Springel_etal08}, which is indicated by the black solid line in Figure~\ref{fig:vmax_rmax}. For the DDM subhalos, both $V_{\mathrm{max}}$ and $R_{\mathrm{max}}$ have changed significantly. If we focus on the ten most massive subhalos (ranked by $V_{\mathrm{max}}$) which are shown with filled circles, we can see that the deviation is more prominent for objects with smaller $V_{\mathrm{max}}$. There is one object in the DDM simulation that has an especially shallow profile. This object has $V_{\mathrm{max}}$=28.21 km/s and $R_{\mathrm{max}}$ = 7.86 kpc in the CDM simulation; however, the decays have caused significant changes in the halo and this object has $V_{\mathrm{max}}$=21.9 km/s and $R_{\mathrm{max}}$=11.9 kpc (the filled orange point near $R_{\mathrm{max}}$ =10 kpc) in the Z13 DDM simulation. The broad behavior of subhalo structure in these DDM models is that, for a given $V_{\mathrm{max}}$, that radius at which this peak velocity is achieved, $R_{\mathrm{max}}$, tends to be larger than in CDM because individual halos of a given mass are more diffuse. 
At a given $V_{\mathrm{max}}$, subhalos in the DDM model tend to have larger values of $R_{\mathrm{max}}$, particularly for $V_{\mathrm{max}} \lsim$ $V_k$  (here 20 km/s), below which the number of subhalos is also dramatically reduced. There are no objects that have $R_{\mathrm{max}} <$ 1 kpc in the DDM simulation in the range of $V_{\mathrm{max}}$ shown in Figure~\ref{fig:vmax_rmax}, while the majority of the small CDM objects scatter around $R_{\mathrm{max}}$ = 1 kpc.
%The discrepancy is especially significant for $V_{\mathrm{max}} \lsim$ 20 km/s ($V_k=20\, \mathrm{km/s}$ in this case), below which the number of subhalos is dramatically reduced. 

\subsubsection{Comparison with the Milky-Way Dwarf Spheroidals}
\label{subsubsection:dSph}

~\cite{Boylan-Kolchin_etal11, Boylan-Kolchin_etal12} highlighted that the inner dynamical structure of the most massive Galactic subhalos in numerical CDM-only simulations is inconsistent with the observed Milky-Way dSph satellites. This finding can be interpreted one of two ways.  First, it could imply that a population of massive subhalos around the Milky Way exists that does not host satellite galaxies.  Another, second, interpretation is that these massive subhalos do host satellites, but that the density profile is dramatically altered in the inner regions relative to CDM.  For the first interpretation, several authors have discussed ways to remove the necessity of this population of massive subhalos.  For example, the Galactic host halo mass, which is usually taken to range over $M_{\mathrm{vir}} \sim$ $(0.6-2.0)\times 10^{12}$ $M_{\odot}$, has been found to be an important factor in determining the abundance of massive subhalos.~\cite{Wang_J_etal12} find that a smaller halo mass $\sim (0.8-1)\times10^{12}M_{\odot}$ will reduce the probability of massive subhalo with $V_{\mathrm{max}} >$ 30 km/s significantly, but the probability of having Magellanic-cloud-like systems is also reduced. There are also questions about the diversities in galactic halo properties and formation history. For example~\cite{Purcell_etal12} show that it is possible that not all systems would have a population of massive subhalos. 

Many authors that have also examined how different DM properties alter the population of massive subhalos, trying to solve the central density problem with altered density profiles instead of deleting massive subhalos. Specific DM models include WDM \citep{Lovell_etal12}, SIDM~\citep{Vogelsberger_etal12}, and mixed DM (warm+cold)~\citep{Anderhalden_etal13}. These models either alter the inner density structure of subhalos (i.e. SIDM) or affect subhalo formation time to make them less centrally concentrated (i.e. WDM). Here we examine the corresponding predictions of DDM. 

In order to reduce the uncertainties due to numerical resolution, we only perform analysis on the high-resolution simulation sets (Z13). As described earlier in ~\S~\ref{subsubsection:dSph}, we correct the effects of force softening by modeling the density distribution using the Einasto fit for $r \leq$ 800 pc and switch back to the simulated density profile for $r >$ 800 pc to ensure robustness on two-body relaxation criterion. Here we only examine one DDM model which is not ruled out by contemporary data and has a significant impact on the structures of Galactic subhalos (DDM with $\Gamma^{-1}$ = 10 Gyr, $V_k$ = 20 km/s). However, we remind the reader that, as discussed previously in~\S~\ref{subsubsection:density_profiles}, there are a range of decay parameters that are capable of reducing subhalo densities and subhalo counts. The main purpose of our present analysis is to establish predictions for the massive subhalo population in DDM models, keeping in mind that the aforementioned issues relating to the lack of precise knowledge of Milky Way DM halo mass and formation history.  
%according to the suggestions in~\cite{Boylan-Kolchin_etal12}.

\begin{figure}
\includegraphics[height=7.5cm]{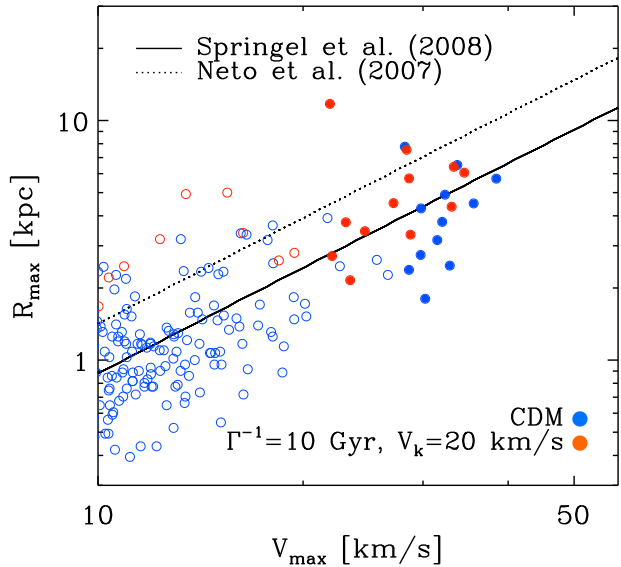} 
\caption{ 
$R_{\mathrm{max}}$ vs. $V_{\mathrm{max}}$ for galactic subhalos from our DDM and CDM zoom-in simulations.  The ten objects with the larges $V_{\mathrm{max}}$ are shown in filled circles. The black solid line shows the galactic subhalo $R_{\mathrm{max}}$ vs. $V_{\mathrm{max}}$ relation from the Aquarius simulation \citep{Springel_etal08} , which shows good agreement with our CDM results. The field halo $R_{\mathrm{max}}$ vs. $V_{\mathrm{max}}$ from \citep{Neto_etal07} is plotted by the black dotted line for reference. 
 }
\label{fig:vmax_rmax}
\end{figure}

\begin{figure*}
\includegraphics[height=8.3cm]{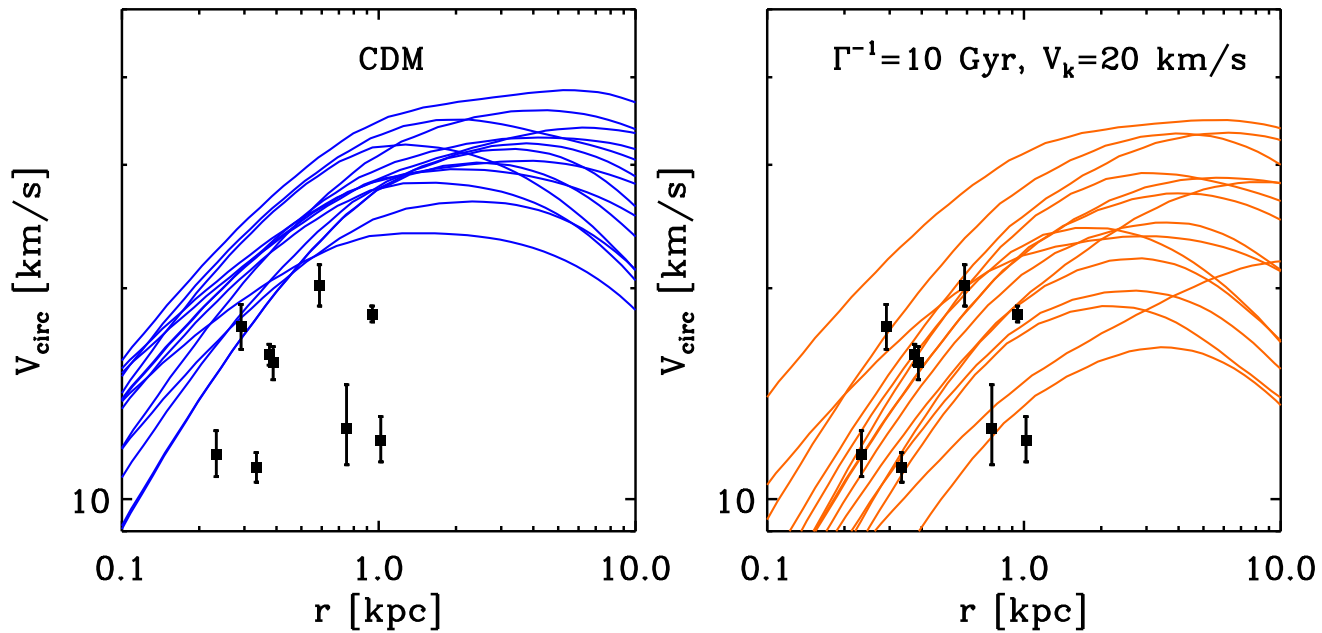}
\caption{ 
Subhalo circular velocity profiles at $z=0$ for the top 15 most massive subhalos (ranked by $V_{\mathrm{max}}$) from our Z13 zoom-in simulations. The left panel show the standard CDM case, and the right panel show the results from DDM model with $\Gamma^{-1}$ = 10 Gyr, $V_k$ = 20 km/s. The square points with 1 $\sigma$ error bars show the observational estimation of the $V_{cir}$ at half-light radius for the 9 classical Milky-Way dSphs \citep{Walker_etal09,Wolf_etal10}. 
}
\label{fig:V_cir}
\end{figure*}

In Figure~\ref{fig:V_cir} we show the circular velocity profiles for the 15 most massive subhalos ranked by $V_{\mathrm{max}}$ at $z=0$ in the Z13 simulations for DDM and CDM models. The left panel shows the standard CDM case, while the right panel is drawn from the DDM simulation with $\Gamma^{-1}$ = 10 Gyr, $V_k$ = 20 km/s. These again are from our high-resolution zoom-in simulations (Z13) after the resolution correction suggested by~\cite{Boylan-Kolchin_etal12}. The square points with 1 $\sigma$ error bars show the observational estimation of the $V_{\mathrm{circ}}$ at the half-light radius for the 9 classical Milky Way dSphs~\citep{Walker_etal09,Wolf_etal10}. This simple exercise demonstrates that DDM  models can provide predictions that are distinct from CDM. Indeed, the circular velocity profiles shown for the DDM model in Figure~\ref{fig:V_cir} certainly appear to be more representative of subhalos that should be broadly consistent with the observed stellar kinematics in the satellite galaxies of the Milky Way.  The circular velocity curves are lower for the DDM model relative to the CDM model both because the number of high-$V_{\mathrm{max}}$ halos is reduced relative to CDM (Fig. \ref{fig:vfn_Z12}) and because the remaining halos are more diffuse than CDM halos (Fig. \ref{fig:vmax_rmax}).  Further, the model that we consider is consistent with current Lyman-$\alpha$ forest constraints, which exclude models with $\Gamma^{-1}$ = 10 Gyr, $V_k \gsim$ 40 km/s \citep{Wang_etal13}, while reducing the abundance of subhalos of a given size relative CDM predictions {\em and} reducing the central densities of subhalos relative to CDM and alleviating the ``too big to fail" problem CDM. 

The Galactic halo we chose to resimulate in this analysis has a mass near the lower end of the range suspected for the Milky Way, and a correspondingly small number of massive subhalos. Our analysis of this halo agrees qualitatively with what is found in \cite{Boylan-Kolchin_etal11} in their analysis of the Aquarius-B or F halo, which have similar halo masses ($M_{\mathrm{vir}}$ $\sim$ 9.54 $\times 10^{11} M_{\odot}$ for Aq-B and $M_{\mathrm{vir}}$ $\sim$ 1.32 $\times 10^{12} M_{\odot}$ for Aq-F). These two halos possess 4 and 5 subhalos that are at least 2 $\sigma$ denser than every bright MW dwarf spheroidal data. It is interesting to note the difference in $\sigma_8$ between the Aquarius simulations ($\sigma_8$ = 0.9) and our simulations ($\sigma_8$ = 0.8). This offset will produce less concentrated subhalos~\citep{zentner_bullock03, Polisensky_etal14} and also reduce the abundance of massive subhalos. However, \citet{Garrison-Kimmel_etal14} find that high subhalo central densities persist for the lower $\sigma_8$ = 0.8, and that those phenomena may continue into the field dwarf halos in our Local Group. As the mass of our simulated host halo is at the low-mass end of the conventional Galactic halo mass range ($M_{\mathrm{vir}}=1\times10^{12}M_{\odot}$), our halo lacks subhalos that can host the LMC or SMC, which are expected to have $V_{\mathrm{max}} >$ 40 km/s. Previous numerical simulations have found that the probability of finding Magellanic Could-like objects can be reduced to $\sim$ 10$\%$ with host halo masses $M_{\mathrm{vir}} \lsim 10^{12}\, \mathrm{M}_{\odot}$ (i.e. \cite{Boylan-Kolchin_etal10,Busha_etal11}, and these results are consistent with observations \citep{Liu_etal11}. Thus our halo is not peculiar in any sense. Previous studies have excluded Magellanic Could-like objects when performing similar analyses because they are many orders of magnitude brighter than the classical dSphs, so our results are quite comparable to the previous literature. %Since our main goal is to understand the ability of DDM model for alleviating the issues in matching dSph data, it is out of the scope of this study to discuss the issues related to host halo mass.  
%Indeed, our initial conditions (and thus our halo final halo virial mass) was chosen to be identical to previous studies of Galactic and sub-Galactic structure {\bf [ALL RELEVANT CITATIONS HERE]} in order to facilitate direct comparisons. 

\section{Discussion and Conclusion}
\label{section:conclusion}

We present a set of N-body simulations of a class of dark matter models in which an unstable dark matter particle decays into a stable dark matter particle and a light non-interacting particle. We especially focus on late-decay scenarios, in which the decay lifetime is comparable to or greater than the Hubble time. We study the effects of the recoil velocities ($V_k$) received by the stable, daughter dark matter particles on the structure dark matter halos with halo mass ranging from galaxy cluster to Milky Way-like objects. We use high-resolution, zoom-in simulations to explore the effects of dark matter decays on Galactic structure and substructure. We explicitly study decaying dark matter models that do not violate current Lyman-$\alpha$ forest constraints~\citep{Wang_etal13}. 

Using a uniform-resolution cosmological simulations within cubic box of side length 50 $h^{-1}$Mpc, we study the large-scale phenomena of DDM models. Here we focused on a DDM model with lifetime $\Gamma^{-1}$ = 40 Gyr, and recoil speed $V_k$ = 100 km/s, which is likely to represent an upper limit on the range of plausible effects DDM may have on halos larger than the Milky Way. We found that for the model with  $\Gamma^{-1}$ = 40 Gyr, $V_k$ = 100 km/s, the influence of DDM is limited to scales $\lsim$ a few kpc. Structure formation in this scenario exhibits strong time evolution such that significant differences between DDM and CDM emerge only at $z \sim$ 1. We find that decay models that exhibit interesting implications on Galactic scales have very small effects on galaxy clusters. 
While we compare the changes in the density profiles and subhalo abundances for halos with different mass by combining with the galactic zoom-in simulation results,

Using Galactic zoom-in simulations we studied the effect of DDM on Milky Way-mass halos and the subhalos of Galaxy-sized systems. On Milky Way-mass scales, we find that DDM models can significantly alter various subhalo properties. Specifically, we find the following results that distinguish DDM models from CDM and other alternative DM models:
%Interesting results that distinguish DDM models from CDM and other alternative DM models include the following.

\textit{Subhalo abundance:}  We find that there is significant impact on Galactic subhalo abundances, in agreement with what is found in \cite{peter_benson10} using semi-analytical models. The amplitude of the change and the range of subhalo sizes over which the change is significant depends on the specific choice of parameters. For $V_k \gsim$ 50 km/s, the effect is a uniform suppression of all galactic subhalos. More interestingly, for lifetimes $\Gamma^{-1} \lsim$ 40 Gyr and $V_k \sim$ 20-40 km/s, DM decays bring the cumulative number of objects with $V_{\mathrm{max}} >$ 10 km/s down to about $\sim 100$. There have been many similar studies on WDM subhalo abunace (i.e. \cite{Polisensky_etal11, Lovell_etal14}), which also find that WDM can efficiently reduce the number of small subhalos. With a thermal relic particle mass $\gsim 2-3$ keV, it is shown that WDM is consistent with the observed number of satellites by SDSS. For mixed DM model where a part of DM consists of WDM and part of it is cold, it is shown in~\cite{Anderhalden_etal13} that the WDM mass limit be lower and still be consistent with observations, since only a portion of the DM is WDM. In contrast, SIDM models exhibit limited power to reduce subhalo abundance \citep{Rocha_etal12}.  

\textit{Inner structure of Milky Way satellites:} We found that DDM models with $V_k \sim$ 10-20 km/s and $\Gamma^{-1} \lsim H_0^{-1}$ yield subhalos with markedly lower internal densities. Indeed, we showed that such a model produces subhalo circular velocities that are broadly consistent with the kinematic constraints on the largest satellite galaxies of the Milky Way. WDM simulations find that subhalos are less concentrated than their CDM counterparts~\citep{Colin_etal00, Lovell_etal14}, mainly because halos of a given mass form later in a WDM cosmology than in CDM, when the mean density of the Universe is lower~\citep{Avila-Reese_etal01,Lovell_etal12}. Recent numerical simulations of viable WDM models~\citep{Maccio_etal12,Shao_etal13} have shown that the density profiles result in a ``soft" core instead of a core large enough that may be relevant for some observations of LSB galaxies~\citep{KuziodeNaray_etal11,Oh_etal11}. Recent cosmological simulations have shown that SIDM is expected to form constant-density isothermal cores from cluster-sized halos down to galactic subhalos \citep{Vogelsberger_etal12,Rocha_etal12}. SIDM models also tend to significantly reduce the central densities of subhalos while leaving the remaining halo structure intact. As we briefly mention in \S~\ref{subsubsection:dSph}, many authors have examined the population of massive subhalos in alternative DM models~\citep{Lovell_etal12, Vogelsberger_etal12, Anderhalden_etal13}. Here we show that DDM models can reduce the central density of the most massive subhalos in a regime of model parameter regime that does not violate the Lyman-$\alpha$ forest limits \citep{Wang_etal13}. Here we only show one particular case, but note that it is likely that several other models have similar interesting phenomenology. An interesting next step involves understanding the properties of subhalos that host Milky Way satellites in different DM scenarios and performing tests on these models.

The preceding points suggest that viable models of decaying dark matter may simultaneously evade constraints on small-scale structure formation from high-redshift observations, such as the Lyman-$\alpha$ forest \citep{Wang_etal13} while alleviating both the ``missing satellites problem" and the ``too big to fail" problem associated with the substructure of Milky Way-sized dark matter halos as predicted by CDM. This is a novel approach to address these discrepancies with an exotic dark matter model that can yield predictions that will be testable with the coming generation of Galactic and extra-galactic surveys.

%------------------------------------------------
%
% acknowledgments
%
%------------------------------------------------

\section*{Acknowledgments}

We thank James Bullock and Michael Boylan-Kolchin for helpful comments and discussions. We also thank Volker Springel for giving us access to Gadget-3. The work of MW and ARZ was supported in part by the Pittsburgh Particle physics, Astrophysics, and Cosmology Center at the University of Pittsburgh and by Grant NSF PHY 0968888 from the U.S. National Science Foundation. AHGP was supported by NASA grant NNX09AD09G and a McCue Fellowship awarded through the Center for Cosmology while at UC Irvine. Simulations were performed on the Frank supercomputer at the Center for Simulation $\&$ Modeling (SaM) at the University of Pittsburgh and on the Big Red 2 supercomputer at Indiana University. This research was supported in part by Lilly Endowment, Inc., through its support for the Indiana University Pervasive Technology Institute, and in part by the Indiana METACyt Initiative. The Indiana METACyt Initiative at IU is also supported in part by Lilly Endowment, Inc.

%------------------------------------------------
% references

\bibliography{simddm}

\begin{thebibliography}{}

\bibitem[\protect\citeauthoryear{{Adams}, {Simon}, {Fabricius}, {van den
  Bosch}, {Barentine}, {Bender}, {Gebhardt}, {Hill}, {Murphy}, {Swaters},
  {Thomas} \& {van de Ven}}{{Adams} et~al.}{2014}]{Adams2014}
{Adams} J.~J.,  {Simon} J.~D.,  {Fabricius} M.~H.,  {van den Bosch} R.~C.~E.,
  {Barentine} J.~C.,  {Bender} R.,  {Gebhardt} K.,  {Hill} G.~J.,  {Murphy}
  J.~D.,  {Swaters} R.~A.,  {Thomas} J.,    {van de Ven} G.,  2014, ArXiv
  e-prints

\bibitem[\protect\citeauthoryear{{Anderhalden}, {Diemand}, {Bertone},
  {Macci{\`o}} \& {Schneider}}{{Anderhalden} et~al.}{2012}]{Anderhalden_etal12}
{Anderhalden} D.,  {Diemand} J.,  {Bertone} G.,  {Macci{\`o}} A.~V.,
  {Schneider} A.,  2012, \jcap, 10, 47

\bibitem[\protect\citeauthoryear{{Anderhalden}, {Schneider}, {Macci{\`o}},
  {Diemand} \& {Bertone}}{{Anderhalden} et~al.}{2013}]{Anderhalden_etal13}
{Anderhalden} D.,  {Schneider} A.,  {Macci{\`o}} A.~V.,  {Diemand} J.,
  {Bertone} G.,  2013, \jcap, 3, 14

\bibitem[\protect\citeauthoryear{{Angulo}, {Hahn} \& {Abel}}{{Angulo}
  et~al.}{2013}]{Angulo_etal13}
{Angulo} R.~E.,  {Hahn} O.,    {Abel} T.,  2013, \mnras, 434, 3337

\bibitem[\protect\citeauthoryear{{Aoyama}, {Sekiguchi}, {Ichiki} \&
  {Sugiyama}}{{Aoyama} et~al.}{2014}]{Aoyama_etal14}
{Aoyama} S.,  {Sekiguchi} T.,  {Ichiki} K.,    {Sugiyama} N.,  2014, ArXiv
  e-prints

\bibitem[\protect\citeauthoryear{{Avila-Reese}, {Col{\'{\i}}n}, {Valenzuela},
  {D'Onghia} \& {Firmani}}{{Avila-Reese} et~al.}{2001}]{Avila-Reese_etal01}
{Avila-Reese} V.,  {Col{\'{\i}}n} P.,  {Valenzuela} O.,  {D'Onghia} E.,
  {Firmani} C.,  2001, \apj, 559, 516

\bibitem[\protect\citeauthoryear{{Benson}, {Frenk}, {Lacey}, {Baugh} \&
  {Cole}}{{Benson} et~al.}{2002}]{benson2002}
{Benson} A.~J.,  {Frenk} C.~S.,  {Lacey} C.~G.,  {Baugh} C.~M.,    {Cole} S.,
  2002, MNRAS, 333, 177

\bibitem[\protect\citeauthoryear{{Blumenthal}, {Faber}, {Primack} \&
  {Rees}}{{Blumenthal} et~al.}{1984}]{Blumenthal_etal84}
{Blumenthal} G.~R.,  {Faber} S.~M.,  {Primack} J.~R.,    {Rees} M.~J.,  1984,
  Nature, 311, 517

\bibitem[\protect\citeauthoryear{{Bode}, {Ostriker} \& {Turok}}{{Bode}
  et~al.}{2001}]{Bode_etal01}
{Bode} P.,  {Ostriker} J.~P.,    {Turok} N.,  2001, \apj, 556, 93

\bibitem[\protect\citeauthoryear{{Boehm}, {Schewtschenko}, {Wilkinson}, {Baugh}
  \& {Pascoli}}{{Boehm} et~al.}{2014}]{Boehm_etal14}
{Boehm} C.,  {Schewtschenko} J.~A.,  {Wilkinson} R.~J.,  {Baugh} C.~M.,
  {Pascoli} S.,  2014, ArXiv e-prints

\bibitem[\protect\citeauthoryear{{Boyarsky}, {Lesgourgues}, {Ruchayskiy} \&
  {Viel}}{{Boyarsky} et~al.}{2009}]{boyarsky_etal08}
{Boyarsky} A.,  {Lesgourgues} J.,  {Ruchayskiy} O.,    {Viel} M.,  2009,
  Journal of Cosmology and Astro-Particle Physics, 5, 12

\bibitem[\protect\citeauthoryear{{Boylan-Kolchin}, {Bullock} \&
  {Kaplinghat}}{{Boylan-Kolchin} et~al.}{2011}]{Boylan-Kolchin_etal11}
{Boylan-Kolchin} M.,  {Bullock} J.~S.,    {Kaplinghat} M.,  2011, \mnras, 415,
  L40

\bibitem[\protect\citeauthoryear{{Boylan-Kolchin}, {Bullock} \&
  {Kaplinghat}}{{Boylan-Kolchin} et~al.}{2012}]{Boylan-Kolchin_etal12}
{Boylan-Kolchin} M.,  {Bullock} J.~S.,    {Kaplinghat} M.,  2012, \mnras, 422,
  1203

\bibitem[\protect\citeauthoryear{{Boylan-Kolchin}, {Springel}, {White} \&
  {Jenkins}}{{Boylan-Kolchin} et~al.}{2010}]{Boylan-Kolchin_etal10}
{Boylan-Kolchin} M.,  {Springel} V.,  {White} S.~D.~M.,    {Jenkins} A.,  2010,
  \mnras, 406, 896

\bibitem[\protect\citeauthoryear{{Brooks}, {Kuhlen}, {Zolotov} \&
  {Hooper}}{{Brooks} et~al.}{2013}]{brooks_etal13}
{Brooks} A.~M.,  {Kuhlen} M.,  {Zolotov} A.,    {Hooper} D.,  2013, \apj, 765,
  22

\bibitem[\protect\citeauthoryear{{Bryan} \& {Norman}}{{Bryan} \&
  {Norman}}{1998}]{bryan1998}
{Bryan} G.~L.,  {Norman} M.~L.,  1998, \apj, 495, 80

\bibitem[\protect\citeauthoryear{{Buckley}, {Zavala}, {Cyr-Racine}, {Sigurdson}
  \& {Vogelsberger}}{{Buckley} et~al.}{2014}]{Buckley_etal14}
{Buckley} M.~R.,  {Zavala} J.,  {Cyr-Racine} F.-Y.,  {Sigurdson} K.,
  {Vogelsberger} M.,  2014, ArXiv e-prints

\bibitem[\protect\citeauthoryear{{Bullock}, {Kravtsov} \& {Weinberg}}{{Bullock}
  et~al.}{2000}]{bullock2000}
{Bullock} J.~S.,  {Kravtsov} A.~V.,    {Weinberg} D.~H.,  2000, \apj, 539, 517

\bibitem[\protect\citeauthoryear{{Busha}, {Wechsler}, {Behroozi}, {Gerke},
  {Klypin} \& {Primack}}{{Busha} et~al.}{2011}]{Busha_etal11}
{Busha} M.~T.,  {Wechsler} R.~H.,  {Behroozi} P.~S.,  {Gerke} B.~F.,  {Klypin}
  A.~A.,    {Primack} J.~R.,  2011, \apj, 743, 117

\bibitem[\protect\citeauthoryear{Cembranos, Feng, Rajaraman \&
  Takayama}{Cembranos et~al.}{2005}]{Cembranos:2005us}
Cembranos J.~A.,  Feng J.~L.,  Rajaraman A.,    Takayama F.,  2005,
  Phys.Rev.Lett., 95, 181301

\bibitem[\protect\citeauthoryear{{Col{\'{\i}}n}, {Avila-Reese} \&
  {Valenzuela}}{{Col{\'{\i}}n} et~al.}{2000}]{Colin_etal00}
{Col{\'{\i}}n} P.,  {Avila-Reese} V.,    {Valenzuela} O.,  2000, \apj, 542, 622

\bibitem[\protect\citeauthoryear{{de Blok} \& {Bosma}}{{de Blok} \&
  {Bosma}}{2002}]{deBlok_etal02}
{de Blok} W.~J.~G.,  {Bosma} A.,  2002, \aap, 385, 816

\bibitem[\protect\citeauthoryear{{Debattista}, {Moore}, {Quinn}, {Kazantzidis},
  {Maas}, {Mayer}, {Read} \& {Stadel}}{{Debattista}
  et~al.}{2008}]{debattista_etal08}
{Debattista} V.~P.,  {Moore} B.,  {Quinn} T.,  {Kazantzidis} S.,  {Maas} R.,
  {Mayer} L.,  {Read} J.,    {Stadel} J.,  2008, \apj, 681, 1076

\bibitem[\protect\citeauthoryear{{Diemand}, {Kuhlen} \& {Madau}}{{Diemand}
  et~al.}{2007}]{VLI}
{Diemand} J.,  {Kuhlen} M.,    {Madau} P.,  2007, \apj, 657, 262

\bibitem[\protect\citeauthoryear{{Garrison-Kimmel}, {Boylan-Kolchin}, {Bullock}
  \& {Kirby}}{{Garrison-Kimmel} et~al.}{2014}]{Garrison-Kimmel_etal14}
{Garrison-Kimmel} S.,  {Boylan-Kolchin} M.,  {Bullock} J.~S.,    {Kirby} E.~N.,
   2014, ArXiv e-prints

\bibitem[\protect\citeauthoryear{{Garrison-Kimmel}, {Boylan-Kolchin}, {Bullock}
  \& {Lee}}{{Garrison-Kimmel} et~al.}{2014}]{ELVIS}
{Garrison-Kimmel} S.,  {Boylan-Kolchin} M.,  {Bullock} J.~S.,    {Lee} K.,
  2014, \mnras, 438, 2578

\bibitem[\protect\citeauthoryear{{Garrison-Kimmel}, {Rocha}, {Boylan-Kolchin},
  {Bullock} \& {Lally}}{{Garrison-Kimmel}
  et~al.}{2013}]{Garrison-Kimmel_etal13}
{Garrison-Kimmel} S.,  {Rocha} M.,  {Boylan-Kolchin} M.,  {Bullock} J.,
  {Lally} J.,  2013, ArXiv e-prints

\bibitem[\protect\citeauthoryear{{Governato}, {Zolotov}, {Pontzen},
  {Christensen}, {Oh}, {Brooks}, {Quinn}, {Shen} \& {Wadsley}}{{Governato}
  et~al.}{2012}]{Governato_etal12}
{Governato} F.,  {Zolotov} A.,  {Pontzen} A.,  {Christensen} C.,  {Oh} S.~H.,
  {Brooks} A.~M.,  {Quinn} T.,  {Shen} S.,    {Wadsley} J.,  2012, \mnras, 422,
  1231

\bibitem[\protect\citeauthoryear{{Hinshaw}, {Larson}, {Komatsu}, {Spergel},
  {Bennett}, {Dunkley}, {Nolta}, {Halpern}, {Hill}, {Odegard}, {Page}, {Smith},
  {Weiland}, {Gold} \& et al.}{{Hinshaw} et~al.}{2012}]{WMAP9}
{Hinshaw} G.,  {Larson} D.,  {Komatsu} E.,  {Spergel} D.~N.,  {Bennett} C.~L.,
  {Dunkley} J.,  {Nolta} M.~R.,  {Halpern} M.,  {Hill} R.~S.,  {Odegard} N.,
  {Page} L.,  {Smith} K.~M.,  {Weiland} J.~L.,  {Gold} B.,    et al. 2012,
  ArXiv e-prints

\bibitem[\protect\citeauthoryear{{Horiuchi}, {Humphrey}, {O{\~n}orbe},
  {Abazajian}, {Kaplinghat} \& {Garrison-Kimmel}}{{Horiuchi}
  et~al.}{2014}]{Horiuchi2014}
{Horiuchi} S.,  {Humphrey} P.~J.,  {O{\~n}orbe} J.,  {Abazajian} K.~N.,
  {Kaplinghat} M.,    {Garrison-Kimmel} S.,  2014, \prd, 89, 025017

\bibitem[\protect\citeauthoryear{{Kaplinghat}}{{Kaplinghat}}{2005}]{Kaplinghat_05}
{Kaplinghat} M.,  2005, \prd, 72, 063510

\bibitem[\protect\citeauthoryear{{Katz} \& {White}}{{Katz} \&
  {White}}{1993}]{katz_white93}
{Katz} N.,  {White} S.~D.~M.,  1993, \apj, 412, 455

\bibitem[\protect\citeauthoryear{{Klypin}, {Kravtsov}, {Valenzuela} \&
  {Prada}}{{Klypin} et~al.}{1999}]{klypin_etal99b}
{Klypin} A.,  {Kravtsov} A.~V.,  {Valenzuela} O.,    {Prada} F.,  1999, \apj,
  522, 82

\bibitem[\protect\citeauthoryear{{Knollmann} \& {Knebe}}{{Knollmann} \&
  {Knebe}}{2009}]{Knollmann_etal09}
{Knollmann} S.~R.,  {Knebe} A.,  2009, \apjs, 182, 608

\bibitem[\protect\citeauthoryear{{Koposov}, {Belokurov}, {Evans}, {Hewett},
  {Irwin}, {Gilmore}, {Zucker}, {Rix}, {Fellhauer}, {Bell} \&
  {Glushkova}}{{Koposov} et~al.}{2008}]{Koposov_etal08}
{Koposov} S.,  {Belokurov} V.,  {Evans} N.~W.,  {Hewett} P.~C.,  {Irwin} M.~J.,
   {Gilmore} G.,  {Zucker} D.~B.,  {Rix} H.-W.,  {Fellhauer} M.,  {Bell} E.~F.,
     {Glushkova} E.~V.,  2008, \apj, 686, 279

\bibitem[\protect\citeauthoryear{{Kuzio de Naray}, {McGaugh} \& {de
  Blok}}{{Kuzio de Naray} et~al.}{2008}]{KuziodeNaray_etal08}
{Kuzio de Naray} R.,  {McGaugh} S.~S.,    {de Blok} W.~J.~G.,  2008, \apj, 676,
  920

\bibitem[\protect\citeauthoryear{{Kuzio de Naray} \& {Spekkens}}{{Kuzio de
  Naray} \& {Spekkens}}{2011}]{KuziodeNaray_etal11}
{Kuzio de Naray} R.,  {Spekkens} K.,  2011, \apjl, 741, L29

\bibitem[\protect\citeauthoryear{{Liu}, {Gerke}, {Wechsler}, {Behroozi} \&
  {Busha}}{{Liu} et~al.}{2011}]{Liu_etal11}
{Liu} L.,  {Gerke} B.~F.,  {Wechsler} R.~H.,  {Behroozi} P.~S.,    {Busha}
  M.~T.,  2011, \apj, 733, 62

\bibitem[\protect\citeauthoryear{{Lovell}, {Eke}, {Frenk}, {Gao}, {Jenkins},
  {Theuns}, {Wang}, {White}, {Boyarsky} \& {Ruchayskiy}}{{Lovell}
  et~al.}{2012}]{Lovell_etal12}
{Lovell} M.~R.,  {Eke} V.,  {Frenk} C.~S.,  {Gao} L.,  {Jenkins} A.,  {Theuns}
  T.,  {Wang} J.,  {White} S.~D.~M.,  {Boyarsky} A.,    {Ruchayskiy} O.,  2012,
  \mnras, 420, 2318

\bibitem[\protect\citeauthoryear{{Lovell}, {Frenk}, {Eke}, {Jenkins}, {Gao} \&
  {Theuns}}{{Lovell} et~al.}{2014}]{Lovell_etal14}
{Lovell} M.~R.,  {Frenk} C.~S.,  {Eke} V.~R.,  {Jenkins} A.,  {Gao} L.,
  {Theuns} T.,  2014, \mnras, 439, 300

\bibitem[\protect\citeauthoryear{{Macci{\`o}}, {Paduroiu}, {Anderhalden},
  {Schneider} \& {Moore}}{{Macci{\`o}} et~al.}{2012}]{Maccio_etal12}
{Macci{\`o}} A.~V.,  {Paduroiu} S.,  {Anderhalden} D.,  {Schneider} A.,
  {Moore} B.,  2012, \mnras, 424, 1105

\bibitem[\protect\citeauthoryear{{Macci{\`o}}, {Ruchayskiy}, {Boyarsky} \&
  {Mu{\~n}oz-Cuartas}}{{Macci{\`o}} et~al.}{2013}]{Maccio_etal12b}
{Macci{\`o}} A.~V.,  {Ruchayskiy} O.,  {Boyarsky} A.,    {Mu{\~n}oz-Cuartas}
  J.~C.,  2013, \mnras, 428, 882

\bibitem[\protect\citeauthoryear{{Moore}, {Ghigna}, {Governato}, {Lake},
  {Quinn}, {Stadel} \& {Tozzi}}{{Moore} et~al.}{1999}]{moore_etal99}
{Moore} B.,  {Ghigna} S.,  {Governato} F.,  {Lake} G.,  {Quinn} T.,  {Stadel}
  J.,    {Tozzi} P.,  1999, \apjl, 524, L19

\bibitem[\protect\citeauthoryear{{Neto}, {Gao}, {Bett}, {Cole}, {Navarro},
  {Frenk}, {White}, {Springel} \& {Jenkins}}{{Neto} et~al.}{2007}]{Neto_etal07}
{Neto} A.~F.,  {Gao} L.,  {Bett} P.,  {Cole} S.,  {Navarro} J.~F.,  {Frenk}
  C.~S.,  {White} S.~D.~M.,  {Springel} V.,    {Jenkins} A.,  2007, ArXiv
  e-prints, 706

\bibitem[\protect\citeauthoryear{{O{\~n}orbe}, {Garrison-Kimmel}, {Maller},
  {Bullock}, {Rocha} \& {Hahn}}{{O{\~n}orbe} et~al.}{2014}]{Onorbe_etal14}
{O{\~n}orbe} J.,  {Garrison-Kimmel} S.,  {Maller} A.~H.,  {Bullock} J.~S.,
  {Rocha} M.,    {Hahn} O.,  2014, \mnras, 437, 1894

\bibitem[\protect\citeauthoryear{{Oh}, {de Blok}, {Brinks}, {Walter} \&
  {Kennicutt} Jr.}{{Oh} et~al.}{2011}]{Oh_etal11}
{Oh} S.-H.,  {de Blok} W.~J.~G.,  {Brinks} E.,  {Walter} F.,    {Kennicutt} Jr.
  R.~C.,  2011, \aj, 141, 193

\bibitem[\protect\citeauthoryear{{Peter}}{{Peter}}{2010}]{peter_10}
{Peter} A.~H.~G.,  2010, \prd, 81, 083511

\bibitem[\protect\citeauthoryear{{Peter} \& {Benson}}{{Peter} \&
  {Benson}}{2010}]{peter_benson10}
{Peter} A.~H.~G.,  {Benson} A.~J.,  2010, \prd, 82, 123521

\bibitem[\protect\citeauthoryear{{Peter}, {Moody} \& {Kamionkowski}}{{Peter}
  et~al.}{2010}]{peter_etal10}
{Peter} A.~H.~G.,  {Moody} C.~E.,    {Kamionkowski} M.,  2010, \prd, 81, 103501

\bibitem[\protect\citeauthoryear{{Planck Collaboration}, {Ade}, {Aghanim},
  {Armitage-Caplan}, {Arnaud}, {Ashdown}, {Atrio-Barandela}, {Aumont},
  {Baccigalupi}, {Banday} \& et al.}{{Planck Collaboration}
  et~al.}{2013}]{Planck_13}
{Planck Collaboration} {Ade} P.~A.~R.,  {Aghanim} N.,  {Armitage-Caplan} C.,
  {Arnaud} M.,  {Ashdown} M.,  {Atrio-Barandela} F.,  {Aumont} J.,
  {Baccigalupi} C.,  {Banday} A.~J.,    et al. 2013, ArXiv e-prints

\bibitem[\protect\citeauthoryear{{Polisensky} \& {Ricotti}}{{Polisensky} \&
  {Ricotti}}{2011}]{Polisensky_etal11}
{Polisensky} E.,  {Ricotti} M.,  2011, \prd, 83, 043506

\bibitem[\protect\citeauthoryear{{Polisensky} \& {Ricotti}}{{Polisensky} \&
  {Ricotti}}{2014}]{Polisensky_etal14}
{Polisensky} E.,  {Ricotti} M.,  2014, \mnras, 437, 2922

\bibitem[\protect\citeauthoryear{{Power}, {Navarro}, {Jenkins}, {Frenk},
  {White}, {Springel}, {Stadel} \& {Quinn}}{{Power}
  et~al.}{2003}]{Power_etal03}
{Power} C.,  {Navarro} J.~F.,  {Jenkins} A.,  {Frenk} C.~S.,  {White} S.~D.~M.,
   {Springel} V.,  {Stadel} J.,    {Quinn} T.,  2003, \mnras, 338, 14

\bibitem[\protect\citeauthoryear{{Purcell} \& {Zentner}}{{Purcell} \&
  {Zentner}}{2012}]{Purcell_etal12}
{Purcell} C.~W.,  {Zentner} A.~R.,  2012, \jcap, 12, 7

\bibitem[\protect\citeauthoryear{{Rocha}, {Peter}, {Bullock}, {Kaplinghat},
  {Garrison-Kimmel}, {O{\~n}orbe} \& {Moustakas}}{{Rocha}
  et~al.}{2013}]{Rocha_etal12}
{Rocha} M.,  {Peter} A.~H.~G.,  {Bullock} J.~S.,  {Kaplinghat} M.,
  {Garrison-Kimmel} S.,  {O{\~n}orbe} J.,    {Moustakas} L.~A.,  2013, \mnras,
  430, 81

\bibitem[\protect\citeauthoryear{{S{\'a}nchez-Salcedo}}{{S{\'a}nchez-Salcedo}}{2003}]{Sanchez-Salcedo_03}
{S{\'a}nchez-Salcedo} F.~J.,  2003, The Astrophysical Journal Letters, 591,
  L107

\bibitem[\protect\citeauthoryear{Sawala, Frenk, Fattahi, Navarro, Bower
  et~al.,}{Sawala et~al.}{2014}]{Sawala:2014baa}
Sawala T.,  Frenk C.~S.,  Fattahi A.,  Navarro J.~F.,  Bower R.~G.,    et~al.,
  2014, ArXiv e-prints

\bibitem[\protect\citeauthoryear{{Schneider}, {Anderhalden}, {Macci{\`o}} \&
  {Diemand}}{{Schneider} et~al.}{2014}]{Schneider2014}
{Schneider} A.,  {Anderhalden} D.,  {Macci{\`o}} A.~V.,    {Diemand} J.,  2014,
  \mnras, 441, L6

\bibitem[\protect\citeauthoryear{{Schultz}, {O{\~n}orbe}, {Abazajian} \&
  {Bullock}}{{Schultz} et~al.}{2014}]{schultz_etal14}
{Schultz} C.,  {O{\~n}orbe} J.,  {Abazajian} K.~N.,    {Bullock} J.~S.,  2014,
  ArXiv e-prints

\bibitem[\protect\citeauthoryear{{Shao}, {Gao}, {Theuns} \& {Frenk}}{{Shao}
  et~al.}{2013}]{Shao_etal13}
{Shao} S.,  {Gao} L.,  {Theuns} T.,    {Frenk} C.~S.,  2013, \mnras, 430, 2346

\bibitem[\protect\citeauthoryear{Shi \& Fuller}{Shi \&
  Fuller}{1999}]{Shi:1998km}
Shi X.-D.,  Fuller G.~M.,  1999, Phys.Rev.Lett., 82, 2832

\bibitem[\protect\citeauthoryear{{Simon}, {Bolatto}, {Leroy}, {Blitz} \&
  {Gates}}{{Simon} et~al.}{2005}]{Simon_etal05}
{Simon} J.~D.,  {Bolatto} A.~D.,  {Leroy} A.,  {Blitz} L.,    {Gates} E.~L.,
  2005, \apj, 621, 757

\bibitem[\protect\citeauthoryear{{Springel}}{{Springel}}{2005}]{Springel_etal05}
{Springel} V.,  2005, \mnras, 364, 1105

\bibitem[\protect\citeauthoryear{{Springel}, {Wang}, {Vogelsberger}, {Ludlow},
  {Jenkins}, {Helmi}, {Navarro}, {Frenk} \& {White}}{{Springel}
  et~al.}{2008}]{Springel_etal08}
{Springel} V.,  {Wang} J.,  {Vogelsberger} M.,  {Ludlow} A.,  {Jenkins} A.,
  {Helmi} A.,  {Navarro} J.~F.,  {Frenk} C.~S.,    {White} S.~D.~M.,  2008,
  \mnras, 391, 1685

\bibitem[\protect\citeauthoryear{{Strigari}, {Bullock}, {Kaplinghat}, {Simon},
  {Geha}, {Willman} \& {Walker}}{{Strigari} et~al.}{2008}]{Strigari_etal08}
{Strigari} L.~E.,  {Bullock} J.~S.,  {Kaplinghat} M.,  {Simon} J.~D.,  {Geha}
  M.,  {Willman} B.,    {Walker} M.~G.,  2008, Nature, 454, 1096

\bibitem[\protect\citeauthoryear{{Strigari}, {Frenk} \& {White}}{{Strigari}
  et~al.}{2010}]{Strigari_etal10}
{Strigari} L.~E.,  {Frenk} C.~S.,    {White} S.~D.~M.,  2010, \mnras, 408, 2364

\bibitem[\protect\citeauthoryear{Strigari, Kaplinghat \& Bullock}{Strigari
  et~al.}{2007}]{Strigari:2006jf}
Strigari L.~E.,  Kaplinghat M.,    Bullock J.~S.,  2007, Phys.Rev., D75, 061303

\bibitem[\protect\citeauthoryear{{Tegmark}, {Eisenstein}, {Strauss} \& {et
  al.}}{{Tegmark} et~al.}{2006}]{Tegmark_etal06}
{Tegmark} M.,  {Eisenstein} D.~J.,  {Strauss} M.~A.,    {et al.} 2006, \prd,
  74, 123507

\bibitem[\protect\citeauthoryear{{Teyssier}, {Pontzen}, {Dubois} \&
  {Read}}{{Teyssier} et~al.}{2013}]{Teyssier_etal13}
{Teyssier} R.,  {Pontzen} A.,  {Dubois} Y.,    {Read} J.~I.,  2013, \mnras,
  429, 3068

\bibitem[\protect\citeauthoryear{{Valluri}, {Debattista}, {Quinn} \&
  {Moore}}{{Valluri} et~al.}{2010}]{valluri_etal10}
{Valluri} M.,  {Debattista} V.~P.,  {Quinn} T.,    {Moore} B.,  2010, \mnras,
  403, 525

\bibitem[\protect\citeauthoryear{{Valluri}, {Debattista}, {Stinson}, {Bailin},
  {Quinn}, {Couchman} \& {Wadsley}}{{Valluri} et~al.}{2013}]{valluri_etal13}
{Valluri} M.,  {Debattista} V.~P.,  {Stinson} G.~S.,  {Bailin} J.,  {Quinn}
  T.~R.,  {Couchman} H.~M.~P.,    {Wadsley} J.,  2013, \apj, 767, 93

\bibitem[\protect\citeauthoryear{{Viel}, {Becker}, {Bolton} \&
  {Haehnelt}}{{Viel} et~al.}{2013}]{Viel_etal13}
{Viel} M.,  {Becker} G.~D.,  {Bolton} J.~S.,    {Haehnelt} M.~G.,  2013, ArXiv
  e-prints

\bibitem[\protect\citeauthoryear{{Villaescusa-Navarro} \&
  {Dalal}}{{Villaescusa-Navarro} \& {Dalal}}{2011}]{Villaescusa-Navarro_etal11}
{Villaescusa-Navarro} F.,  {Dalal} N.,  2011, \jcap, 3, 24

\bibitem[\protect\citeauthoryear{{Vogelsberger}, {Zavala} \&
  {Loeb}}{{Vogelsberger} et~al.}{2012}]{Vogelsberger_etal12}
{Vogelsberger} M.,  {Zavala} J.,    {Loeb} A.,  2012, \mnras, 423, 3740

\bibitem[\protect\citeauthoryear{{Walker}, {Mateo}, {Olszewski},
  {Pe{\~n}arrubia}, {Wyn Evans} \& {Gilmore}}{{Walker}
  et~al.}{2009}]{Walker_etal09}
{Walker} M.~G.,  {Mateo} M.,  {Olszewski} E.~W.,  {Pe{\~n}arrubia} J.,  {Wyn
  Evans} N.,    {Gilmore} G.,  2009, \apj, 704, 1274

\bibitem[\protect\citeauthoryear{{Wang}, {Frenk}, {Navarro}, {Gao} \&
  {Sawala}}{{Wang} et~al.}{2012}]{Wang_J_etal12}
{Wang} J.,  {Frenk} C.~S.,  {Navarro} J.~F.,  {Gao} L.,    {Sawala} T.,  2012,
  \mnras, 424, 2715

\bibitem[\protect\citeauthoryear{{Wang} \& {White}}{{Wang} \&
  {White}}{2007}]{Wang_etal07}
{Wang} J.,  {White} S.~D.~M.,  2007, \mnras, 380, 93

\bibitem[\protect\citeauthoryear{{Wang}, {Croft}, {Peter}, {Zentner} \&
  {Purcell}}{{Wang} et~al.}{2013}]{Wang_etal13}
{Wang} M.-Y.,  {Croft} R.~A.~C.,  {Peter} A.~H.~G.,  {Zentner} A.~R.,
  {Purcell} C.~W.,  2013, \prd, 88, 123515

\bibitem[\protect\citeauthoryear{{Wang} \& {Zentner}}{{Wang} \&
  {Zentner}}{2010}]{wang_zentner10}
{Wang} M.-Y.,  {Zentner} A.~R.,  2010, \prd, 82, 123507

\bibitem[\protect\citeauthoryear{{Wang} \& {Zentner}}{{Wang} \&
  {Zentner}}{2012}]{wang_zentner12}
{Wang} M.-Y.,  {Zentner} A.~R.,  2012, \prd, 85, 043514

\bibitem[\protect\citeauthoryear{{White} \& {Rees}}{{White} \&
  {Rees}}{1978}]{White_etal78}
{White} S.~D.~M.,  {Rees} M.~J.,  1978, \mnras, 183, 341

\bibitem[\protect\citeauthoryear{{Wolf}, {Martinez}, {Bullock}, {Kaplinghat},
  {Geha}, {Mu{\~n}oz}, {Simon} \& {Avedo}}{{Wolf} et~al.}{2010}]{Wolf_etal10}
{Wolf} J.,  {Martinez} G.~D.,  {Bullock} J.~S.,  {Kaplinghat} M.,  {Geha} M.,
  {Mu{\~n}oz} R.~R.,  {Simon} J.~D.,    {Avedo} F.~F.,  2010, \mnras, 406, 1220

\bibitem[\protect\citeauthoryear{{Yue} \& {Chen}}{{Yue} \&
  {Chen}}{2012}]{yue_chen12}
{Yue} B.,  {Chen} X.,  2012, \apj, 747, 127

\bibitem[\protect\citeauthoryear{{Zentner} \& {Bullock}}{{Zentner} \&
  {Bullock}}{2003}]{zentner_bullock03}
{Zentner} A.~R.,  {Bullock} J.~S.,  2003, \apj, 598, 49

\bibitem[\protect\citeauthoryear{Zurek}{Zurek}{2014}]{Zurek:2013wia}
Zurek K.~M.,  2014, Phys.Rept., 537, 91

\end{thebibliography}

%------------------------------------------------

%------------------------------------------------
\end{document}